\documentclass[fleqn,usenatbib]{mnras}

\usepackage{newtxtext,newtxmath}

\usepackage[T1]{fontenc}
\usepackage{ae,aecompl}

\usepackage{graphicx}
\usepackage{amsmath}
\usepackage{amssymb}

\title[Assembly Bias and Redshift Space Distortions]{The Effect of Assembly Bias on Redshift Space Distortions}

\author[Padilla et al.]{
N. Padilla$^{1,2}$\thanks{E-mail: npadilla@astro.puc.cl},
S. Contreras$^{1,3,4}$,
I. Zehavi$^{5}$,
C. M. Baugh$^{6}$,
P. Norberg$^{6,7}$
\\
 $^{1}$Instit\'uto Astrof\'{\i}sica, Pontifica
Universidad Cat\'olica de Chile, Santiago, Chile \\
 $^{2}$Centro de Astro-Ingenier\'{\i}a, Pontificia Universidad Cat\'olica de 
Chile, Santiago, Chile \\
 $^{3}$Centro de Estudios de F\'{\i}sica del Cosmos de Arag\'on (CEFCA), 
Plaza San Juan 1, Planta-2, Teruel, E-44001, Spain \\
$^{4}$Donostia International Physics Center (DIPC),
Manuel Lardizabal pasealekua 4, 20018 Donostia, Basque Country, Spain \\
 $^{5}$Department of Physics, Case Western Reserve University, Cleveland, OH
44106, USA \\  
 $^{6}$Institute for Computational Cosmology, Department of Physics, 
Durham University, South Road, Durham, DH1 3LE, UK \\
 $^{7}$Centre for Extragalactic Astronomy, Department of Physics, 
Durham University, South Road, Durham, DH1 3LE, UK \\
}

\date{Accepted XXX. Received YYY; in original form ZZZ}

\pubyear{2018}

\begin{document}
\label{firstpage}
\pagerange{\pageref{firstpage}--\pageref{lastpage}}
\maketitle
%
\begin{abstract}
We study potential systematic effects of assembly bias on cosmological parameter constraints from redshift space distortion measurements.  We use a semi-analytic galaxy formation model applied to the Millennium N-body WMAP-7 simulation to study the effects of halo assembly bias on the redshift space distortions of the galaxy correlation function.  We look at the pairwise velocities of galaxies living in haloes with concentrations and ages in the upper and lower quintiles, and find that the velocity differences between these are consistent with those reported for real-space clustering analyses, i.e. samples with higher clustering also exhibit stronger infall pairwise motions.  This can also be seen  in the monopole and quadrupole of the redshift-space correlation function. We find that regardless of the method of measurement, the changes in the $\beta$ parameter due to different secondary halo parameters fully tracks the change in the bias Parameter.  Hence, assembly bias does not introduce detectable systematics in the inferred logarithmic growth factor.
\end{abstract}


\section{Introduction}

One of the most successful tools to determine the cosmological 
parameters of the Universe 
is the analysis of the galaxy correlation function measured in redshift space 
\citep{Hu:2003,Wagner:2008,Guzzo:2008,Shoji:2009}.  This correlation function differs from 
the real-space correlation function since the inferred comoving positions of galaxies are shifted 
by their peculiar velocities in the direction of the line of sight  
which introduces an anisotropy in the correlation
function. Its strength depends on the
amplitude of peculiar velocities which in turn encodes cosmological information;
the rate and acceleration of the expansion of 
the Universe oppose gravitational collapse and this is encoded in the rate of growth of perturbations 
which is revealed by the peculiar motions of galaxies
\citep{Peebles:1980}.

In some cases the full two-dimensional correlation function $\xi(\mu,s)$ is used to make 
constraints on cosmological parameters, where the dimensions correspond to the pair separation 
in redshift space, $s$, and the cosine of the angle to the line of sight, $\mu$ 
\citep{Okumura:2008,Chuang:2012}.  However, the covariance matrices needed to estimate errors in the parameters
are too large owing to the large number of bins of this 2-dimensional space.  This makes the covariance matrix subject to important uncertainties.  \cite{Padmanabhan:2008} proposed to use 1-dimensional
multipole functions obtained using Legendre polynomials to alleviate this problem and,
at the same time, to
avoid assuming knowledge of the shape of redshift-space distortions; 
they show that the joint analysis of the monopole
and quadrupole provides measurements of the angular diameter distance $D_A(z)$ and the Hubble constant 
$H(z)$ (see also the ``clustering wedges" method proposed by \citealt{Kazin:2012} that constrains 
the same quantities).

When inferring cosmological parameters from redshift space distortions it
is commonly assumed that galaxy velocities are unbiased with respect
to those of the underlying matter.  On small scales there is evidence that
this is not necessarily a good approximation. \cite{Guo:2015a} studied the CMASS
BOSS sample and found that central galaxies show a velocity dispersion of $\sim 20\%$
the total velocity dispersion of the haloes, and that the velocity dispersion of satellites is $\sim 80\%$
of the total.  Furthermore, \cite{Ye:2017} reach similar conclusions for 
the SDSS Main sample analysing jointly with galaxies from the Illustris simulation.
On slightly larger scales, \cite{Hearin:2015} show that the halo assembly 
history, in addition to halo mass, affects the  galaxy
peculiar velocities by about $20\%$ in their amplitude 
up to scales of $10\ {h}^{-1}$Mpc, implying that models 
of the galaxy distribution that intend to use the dynamics
of galaxies on these scales need to take this effect into account.  
On scales corresponding
to galaxies hosted by different haloes (the 2-halo term), it is generally assumed that 
galaxy peculiar velocities are unbiased.  
This appears to be a good approximation; a study of the velocity bias $b_v$ in a cosmological numerical simulation by \cite{Chen:2018} 
shows that there are significant deviations from $b_v=1$ that
depend on redshift and mass, but these are of the order of $1\%$ out to $k=0.2$ h\ Mpc$^{-1}$ at $z=0$,
and can reach $5\%$ on the same scales at higher redshifts $z\sim 2$.

Halo clustering was initially assumed to depend on halo mass alone.  However, 
\cite{Gao:2005} showed that it also depends on halo assembly history (this was also 
noticed in \citealt{Sheth:2004}).
\cite{Croton:2007} termed this effect
assembly bias and demonstrated that it also impacts galaxy clustering.  
The definition of halo assembly bias was expanded to encapsulate the dependence of 
(large-scale) clustering amplitude on other halo properties in addition to halo mass, i.e. a 
second parameter such as spin, concentration or the number of substructures per halo (see e.g. \citealt{Gao:2007}, and \citealt{Mao:2017} for a recent compilation of properties that produce halo assembly bias).

Occupancy variation is the dependence of the halo occupation distribution (i.e. the number of galaxies per halo as a function of halo mass, HOD from this point on) on halo properties in addition to mass \citep{Zehavi:2018}.  This in turn can induce changes in the observed clustering of the galaxies that populate these haloes  (\citealt{Zehavi:2018}; see also \citealt{Artale:2018}, \citealt{C18}).
Both assembly bias and occupancy variation affect the amplitude of clustering of a given sample of galaxies, and their combined effect is commonly referred to as galaxy assembly bias.

Despite several attempts to find hard evidence of assembly bias in galaxy
surveys, it is still not clear whether the different signatures pointed out by
various authors are due to systematic effects that can mimic the effects of assembly bias (e.g., \citealt{Campbell:2015b,Zu:2016a,Busch:2017,Sin:2017,Tinker:2018,Lacerna:2018}).
To date, not only is it still not proven that
 galaxy assembly bias is present in the real Universe, it is also not clear if it 
would have an impact on cosmological constraints, as shown, for
instance, by \cite{Mcewen:2018}.  They show that the cross
correlation between galaxies and mass
shows no effect of assembly bias in
scales $\sim 10\ ${h}$^{-1}$Mpc.  If this extends to the scales usually applied to cosmological 
constraints then this statistic would provide
an avenue for results free from systematics via measurements of galaxy-weak lensing shear statistics.

If galaxy velocities are indeed unbiased in the 2-halo term regime, then the
assembly bias effect seen on spatial clustering statistics should translate completely 
to phase space statistics and thus not introduce any systematic effects in 
cosmological parameter constraints.   This is what we intend to look into here.  To do so we  use semi-analytic galaxies from a numerical simulation
and find out whether velocity differences of galaxies living in haloes of
equal mass but with different secondary parameters can be completely explained by 
differences in their bias parameter as inferred from spatial statistics (i.e. in real space).

Our work complements recent results by \cite{Xu:2018} who look at the change of pairwise velocities of dark-matter haloes as a function of mass and several
secondary parameters.  They find that the changes in pairwise velocities and clustering due to secondary parameters are correlated but that the scatter in
this relation is wide.  Here we 
focus on galaxy populations from a semi-analytic
model and look at the effect of halo concentration and age.
In addition we will also look at the effect of galaxy selection and evolution
with redshift.  Finally, rather than looking at the correlation between changes
in velocity and clustering, we will focus on whether cosmological parameters
obtained from redshit-space distortions could be biased because of
galaxy assembly bias.

This work is the latest installment in a series of papers focusing
on galaxy assembly bias.  In \cite{Zehavi:2018} we made a careful analysis
of the variation of the number of galaxies in haloes due
to different secondary properties and how this combines with halo assembly bias to change the clustering of galaxies.  The next paper
in this series was \cite{C18}, where we looked at the evolution of galaxy
assembly bias with redshift.  In the present installment, we extend the analysis in these
two previous works to velocity space, with special emphasis on redshift-space
distortions.

This paper is organised as follows.  In Section \ref{Sec:Sims} we present the
N-body simulation, the semi-analytic model that is applied to it, and the sample 
selection. In Section \ref{Sec:Shuffle} we present the method to create
shuffled samples of galaxies that remove the occupancy variation but
retain the inner halo structure both in position and velocity space.
Section \ref{Sec:Vcorr} presents results of the pairwise velocities and 
their dependence on parameters other than the halo mass. We extend
the analysis to correlation functions in redshift space in Section \ref{Sec:Mult},
and Section \ref{Sec:Cosm} shows the cosmological parameters one would estimate
from these correlation functions.  Finally we present our conclusions in Section \ref{Sec:End}.  The Appendix
contains additional tests and different ways to calculate the effect of galaxy assembly
bias on the $\beta$ parameter.

\section{Simulations and Samples}
\label{Sec:Sims}

We use numerical simulations populated with a semi-analytic model of galaxy formation.
Semi-analytical models (SAMs) of galaxy formation couple the output of large dark-matter only simulations with simple descriptions of the physical processes affecting the baryonic content of galaxies to help study galaxy formation on a cosmological footing.  SAMs follow a galaxy population as 
it evolves, which allows us to study galaxy samples at different redshifts, selected
according to different physical properties.  We outline the details of the simulation and
the selection of samples below.

\subsection{Semi-analytic model and numerical simulation}

The typical processes that are followed in semi-analytic models are: the collapse and merging of dark matter haloes, shock heating and radiative cooling of gas, star formation, supernovae, AGN, and photoionization feedback, chemical enrichment of  the gas and stars, instabilities of the gaseous disc and galaxy mergers.  

We will focus our analysis on the \cite{Guo:2013a} (hereafter G13) model, which is a version of {\tt L-GALAXIES}, the SAM code of the Munich group \citep{DeLucia:2004,Croton:2006,DeLucia:2007,Guo:2011,Henriques:2013,Henriques:2015}.  An in depth description of the model can be found in Guo et al. (2013, see also \citealt{C17} and \citealt{Guo:2016b}).  

We use outputs obtained after this model is applied to the Millennium-WMAP7 simulation (hereafter MS7;  \citealt{Guo:2013a}), which are publicly  available.\footnote{\url{http://gavo.mpa-garching.mpg.de/Millennium/}} This simulation shares several characteristics with the Millennium simulation \citep{Springel:2005}, but is run using  a WMAP7  cosmology\footnote{The cosmological parameters used in the simulation are  $\Omega_{\rm m0}$= $\Omega_{\rm dm0}$+$\Omega_{\rm b0}$ = 0.272, $\Omega_{\Lambda0}$ = 0.728, $\Omega_{\rm b0}$ = 0.0455, $\sigma_8$ = 0.81, $n_{\rm s}$ = 0.967, ${h}$ = 0.704.}. The simulation follows $2160^3$ particles in a $(500\, {h}^{-1}\, {\rm~Mpc})^3$ periodic comoving box, with a particle mass of $9.31\times10^8 {h}^{-1} \,{\rm M_{\odot}}$. In total, 61 simulation  outputs were stored between $z = 50$ and $z=0$. 

The simulation outputs are used to construct merger trees, which are an input to the SAM.  The haloes are identified using a friends-of-friends ({\tt FoF}) finding algorithm \citep{Davis:1985}.  This is run on each output of the simulation and all haloes with at least 20 particles are saved. {\tt SUBFIND} is then run on these halo catalogues to identify their substructures, called subhaloes  \citep{Springel:2001}.  The final step is to use these subhaloes to build the merger trees by linking each subhalo in one snapshot to a single descendant subhalo in the following output. The mass of the halo ($M_{\rm halo}$) is defined as the mass within the radius where the halo overdensity is 200 times the critical density (referred to as ``m\_crit200'' in the public database).

Galaxies in the \cite{Guo:2013a} model are assigned velocities following the approach set out in \cite{Guo:2011}.  While 
galaxies are associated with a dark matter subhalo (either central or satellite) they are assigned the subhalo velocity until the baryonic mass of the galaxy becomes larger than the dark matter mass of the subhalo.  This happens to subhaloes due to tidal stripping.  This process has a stronger effect on 
extended dark matter haloes compared to the more concentrated baryons in galaxies, which are also
subject to stripping processes followed analytically by the model.
Because the orbit of the subhalo is not accurate after this limit, it is replaced with
that of the most-bound particle of the subhalo, modified by a time-dependent orbit-shrinking factor which models the orbital decay caused by the dynamical friction.  This treatment of satellite velocities produces density profiles of clusters of galaxies in good agreement with observations, but
it is analytic and subject to several assumptions. We will bear this in mind when analysing the redshift space distortions measured from the model output.

\subsection{Classifying by halo properties}

Assembly bias in haloes appears in cosmological simulations when studying samples of equal mass and different secondary property.  We will concentrate on two different secondary properties, the halo formation time with which we separate our galaxies into those living in late and early formed haloes (low and high formation redshift), and the halo concentration to define low and high concentration samples. 

We adopt the redshift at which the halo reaches half its final mass as its formation time, interpolating between  the available simulation snapshots (e.g., \citealt{Gao:2004,Gao:2005,Croton:2007,Zehavi:2018}).  As it evolves, a halo sometimes shows spurious changes of mass  \citep{C17b}; to avoid being affected by this problem we set the formation time as the first occurrence of a halo reaching half its final mass. We calculate the formation redshift at each snapshot; therefore, it is expected that as a halo evolves, its formation time defined in this way also changes.

Halo concentration is defined using the ratio between $V_{\rm max}$ and $V_{\rm vir}$ \citep{Gao:2007}, where $V_{\rm max}$ is the peak value of its circular velocity curve ($V_{\rm c} = (G\ M(r)/r)^{1/2}$) and $V_{\rm vir}$ the virial velocity of a halo ($V_{\rm vir} \equiv V_{\rm c}(r_{\rm vir})$)
\citep{Bullock:2001,Gao:2007}. This definition has the advantage that it can be 
calculated directly with the information available in the public database without the need for model fitting.
Halo concentration can show a strong halo assembly bias effect and is advantageous with respect to age as it can be obtained from single snapshots rather than merger trees (e.g. \citealt{Gao:2005}, \citealt{Wechsler:2006},\citealt{C18}).  
In addition, there are methods to estimate halo concentration in observations (e.g. \citealt{Biviano:2017}). 
Because of this we will pay special attention to the results using
this secondary parameter.

To define our samples of early and late formation haloes (and of high and low concentration) we take the upper and lower quintile of the samples in bins of halo mass of $0.1$ dex, effectively forcing the mass function to be the same in each sample. We test whether the bin size affects studies of assembly bias in \cite{Zehavi:2018} and find little changes for formation time in $z=0$ results. \cite{C18} show that halo age is reliably measured in the Millennium WMAP-7 simulation up to $z=3$.

\begin{figure}
\includegraphics[width=0.49\textwidth]{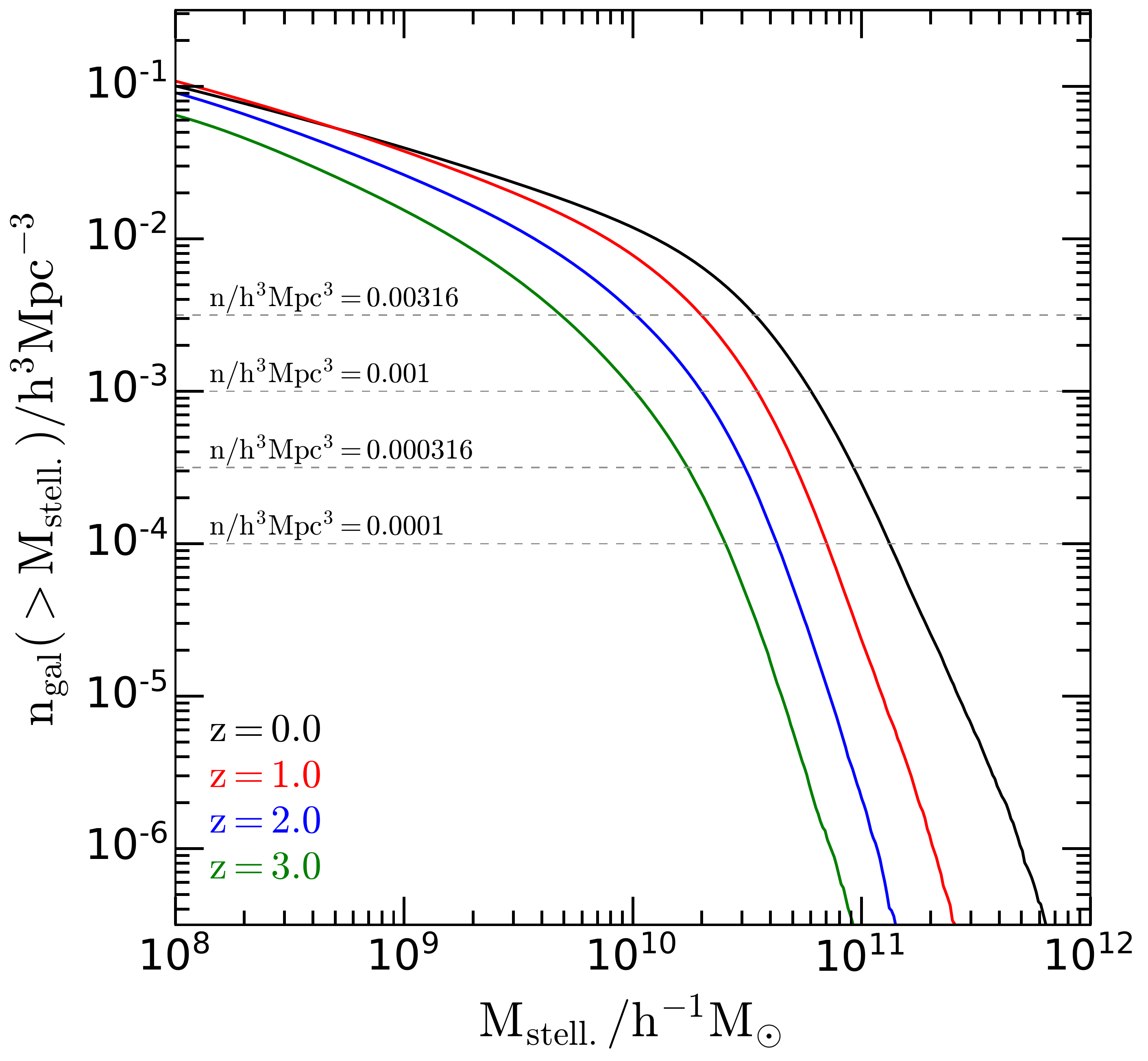}
\includegraphics[width=0.47\textwidth]{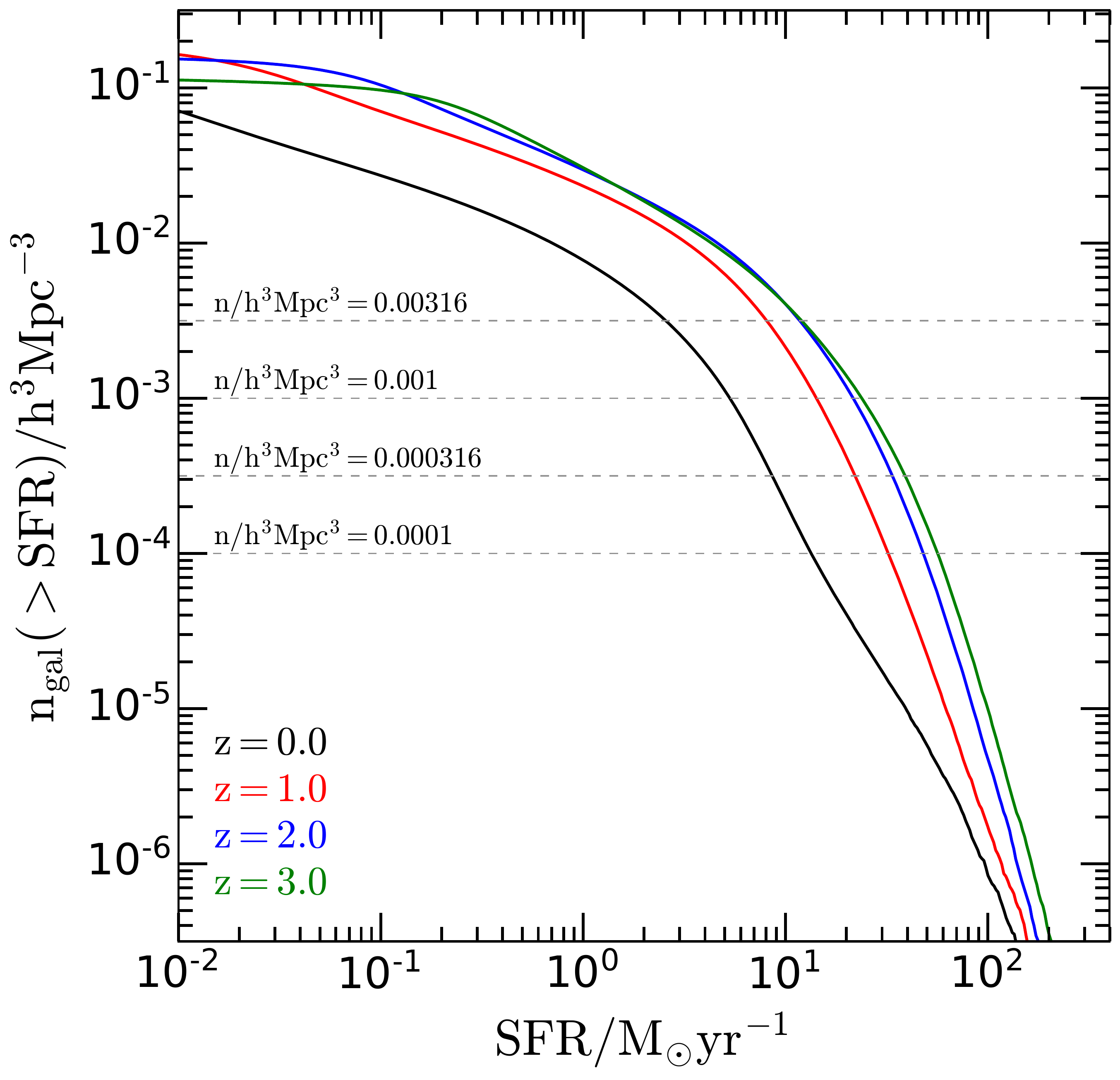}
\caption{Cumulative stellar mass and SFR functions (top and bottom panels, respectively) at $z=0,1,2$ and  $3$ (different colours indicated in the key). The dashed horizontal lines represent the number densities used to select samples of galaxies in this work as labeled.}
\label{Fig:AMF}
\end{figure}

\subsection{Galaxy selection}

We use the number density of galaxies ranked in order of decreasing stellar mass and star formation rate (SFR) to construct galaxy samples from the \cite{Guo:2013a} model.  We will use the following number densities, $n = 0.0316,\ 0.01, \  0.00316,\ 0.001$ and $ 0.000316 \,{h}^{3}\,  {\rm Mpc}^{-3}$ and redshifts $z=0, 1, 2$ and $3$. It is quite common in the literature to use number density to define samples of galaxies from surveys, and in cases where passive evolution can be safely assumed (eg. \citealt{Seo:2008}), to connect galaxy populations at different redshifts \citep{Padilla:2010,Leja:2013,Mundy:2015,Torrey:2015,C17}. A number density selection has the definite advantage of being insensitive to systematics in the estimation of the galaxy property that is selected to rank the galaxies, and can be applied both to simulations and observations which facilitates meaningful comparisons between them.

A stellar mass selection is similar to the selection used in optical or near-infrared surveys, whereas SFR is relevant to observational samples selected by emission line strength or rest-frame UV luminosity.
The samples were chosen to be evenly spaced in logarithmic number density, corresponding to a change of half a decade in log  abundance.

We will place special emphasis on the stellar mass selected samples that mimic the  SDSS main galaxy sample \citep{Strauss:2002}, with $z=0$ and $n=0.001$ ${h}^3$Mpc$^{-3}$, the luminous red galaxies  of SDSS-III and IV (BOSS, \citealt{Eisenstein:2011}, and eBOSS, \citealt{Dawson:2016}) with  $z\sim 1.0$ and $n=0.00031\ {h}^3$Mpc$^{-3}$, and also on  SFR selected samples similar to the eBOSS emission line galaxy sample (ELG, \citealt{Comparat:2016}) corresponding to $z\sim 1.0$ and $n=0.000316\ {h}^3$Mpc$^{-3}$, and to HETDEX \citep{Adams:2011} with $z\sim 2.0$ and  $n=0.0001\ {h}^3$Mpc$^{-3}$.  To illustrate the procedure by which we select our samples we show the cumulative stellar mass and SFR functions in the top
and bottom panels of Fig.~\ref{Fig:AMF} for  all redshifts studied. 
The horizontal dashed lines show the number density cuts that will be used throughout. For a given number density, samples of galaxies are constructed using masses or SFRs higher than their value at the intersection with the associated dashed line.

In order to analyse the clustering of galaxies in redshift space, we use the $Z$-coordinate
of the peculiar velocities to displace the $Z$-axis component of a galaxy position, $Z_s=Z+v_Z/H(z)$, where the subscript $s$ denotes a position in redshift space.

\section{Shuffled catalogues with velocity information}
\label{Sec:Shuffle}

Statistics of the phase space density fields of galaxies, such as galaxy clustering in redshift
space,  should in principle respond to
two effects, halo assembly bias on the one hand, and occupancy variation on the other.  
The halo properties that produce halo assembly bias, i.e. a dependence of 
halo clustering on secondary properties in addition to their mass,
include halo age, spin, number of substructures, concentration, among others (e.g. \citealt{Gao:2007}).
The effect of occupancy variation also changes the amplitude of galaxy
clustering, as haloes of different
secondary properties are populated by different numbers of galaxies; both effects combined
give rise to galaxy assembly bias.

The process of shuffling galaxies between haloes in simulations effectively removes the HOD dependence 
on halo properties, i.e. it removes the occupancy variation component of galaxy 
assembly bias.  This technique
was originally proposed by \cite{Croton:2007} and consists of interchanging the galaxy population of individual haloes in narrow ranges of halo mass, thus removing the connection to the assembly history
of the haloes.  This way haloes of similar mass and different properties do not show differences in the number of galaxies that populate them.  In this work we adopt a halo mass bin of $0.1$dex in ${h}^{-1}M_\odot$.  
Central and satellite galaxies in individual haloes are moved together to a different one while retaining their positions relative to the
centre of potential of the halo, where the central galaxy of a halo is placed.  

An alternative shuffling procedure has been proposed in the literature with the satellites shuffled as well among different halos of the same mass independently of the central galaxy population (e.g., \citealt{Zu:2008,Zentner:2014}). This shuffling is important if one is concerned with the relation between centrals and satellites, or between satellites in the same halo, such as in the case of satellite-central alignments.  In our case this is not needed as 
we will treat centrals and satellites simply as
tracers of the density field.

Here we extend the shuffling procedure to velocity space, where before being assigned to a new halo, the velocities of halo member galaxies are expressed
relative to the velocity of the centre of potential of the haloes.
This process retains the relative velocities between galaxies in different haloes such
as the Kaiser effect but removes, for instance, the dependence of velocity dispersion
on secondary halo parameters, as well as the possible alignments between the velocity and position ellipsoids of satellite systems in neighbouring haloes \citep{FR:2014}.

This shuffling method retains the phase space structure of haloes, and allows us to extend assembly
bias studies to velocity space.  Namely we will look into whether assembly bias affects velocities
in a different way than positions.  Such a difference could lead to potential systematic biases in cosmological
parameter constraints obtained from redshift-space distortion analyses.

\section{Assembly bias and velocity statistics}
\label{Sec:Vcorr}

In this section we present measurements of three-dimensional average radial velocities between pairs of galaxies.  We refer to this statistic as pairwise velocities; these velocities are the ones responsible for the distortion of the correlation function in the direction parallel to the line of sight, as they change the relative separation of pairs along this coordinate when galaxy distances are obtained via their redshift \citep{Hamilton:1992}.  We will measure how the pairwise velocities change for different halo secondary properties, and analyse what is the effect that assembly bias plays in these measurements.

\subsection{Pairwise velocities}

We will show measurements of
pairwise velocities between the galaxies in our samples in different concentration (halo age) ranges, and the full sample. This will allow us to 
maximise the signal-to-noise of our measurements.

For the pairwise velocities we calculate
\begin{equation}
v_{12}(r)=\left< (\bf{v}_i-{\bf v}_j)  \hat{\bf{r}_{ij}} \right> |_r,
\end{equation}
where $r=|\bf{r}_i-\bf{r}_j|$, $\hat{\bf{r}_{ij}}=(\bf{r}_i-\bf{r}_j)/r$, $i$ and $j$ refer to a pair of galaxies, $\bf v$ and $\bf r$ are their comoving velocity and position vectors, and the average is done on pairs separated by distance $r=|{\bf r}_i-{\bf r}_j|$ \citep{Peebles:1980,Fisher:1994}.  In the case of pairwise velocities for one of the high or low concentration (halo age) samples, the index $i$ corresponds to galaxies in one of the quintiles of the secondary halo property, 
and the $j$ index to the full galaxy sample corresponding to the chosen
redshift and space density cut.
As mentioned earlier, we will mostly focus on 
concentration as the second parameter.
Results obtained using age will be shown for some interesting cases in the appendix. 

\begin{figure*}
\includegraphics[width=0.47\textwidth]{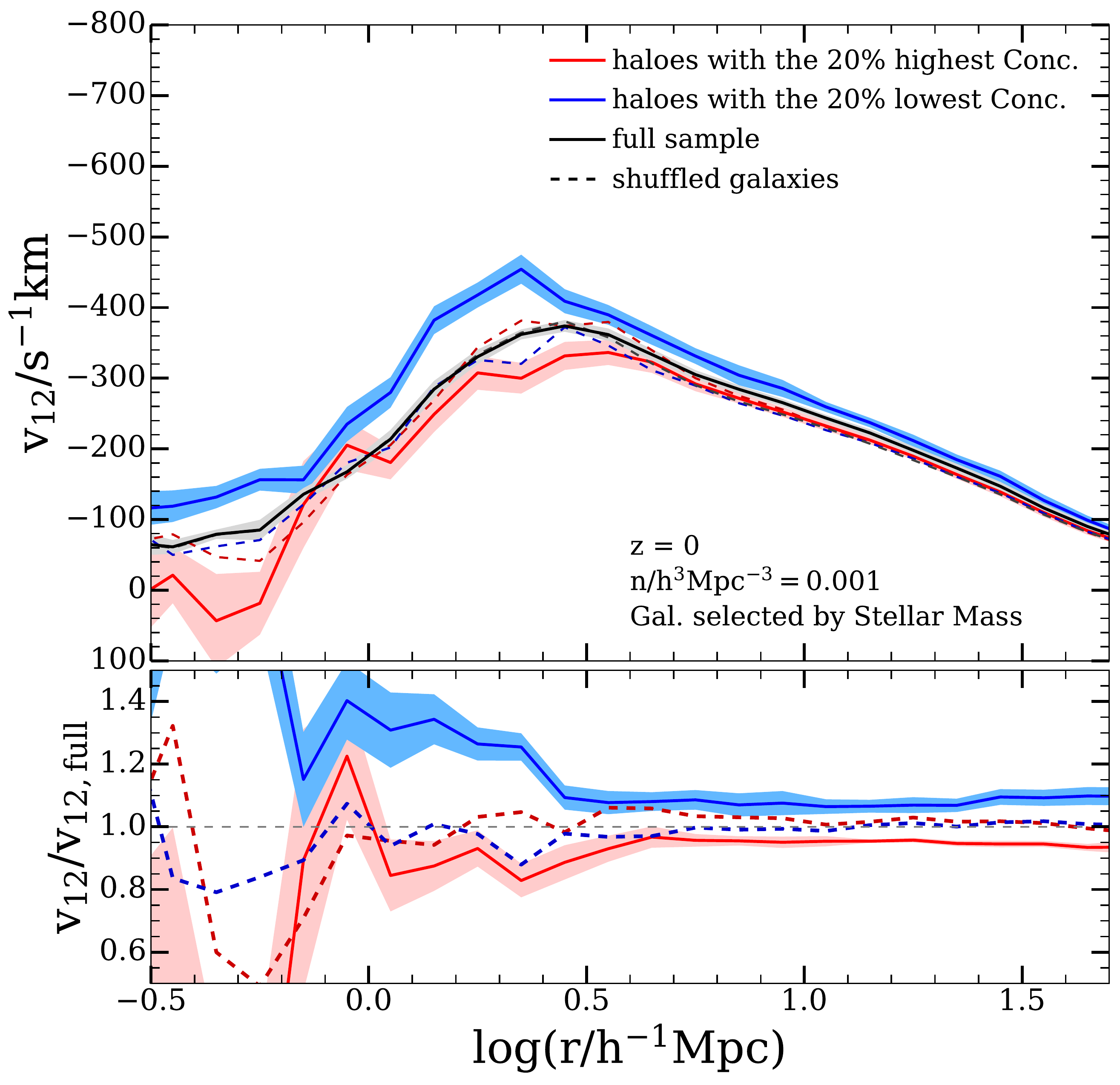}
\includegraphics[width=0.47\textwidth]{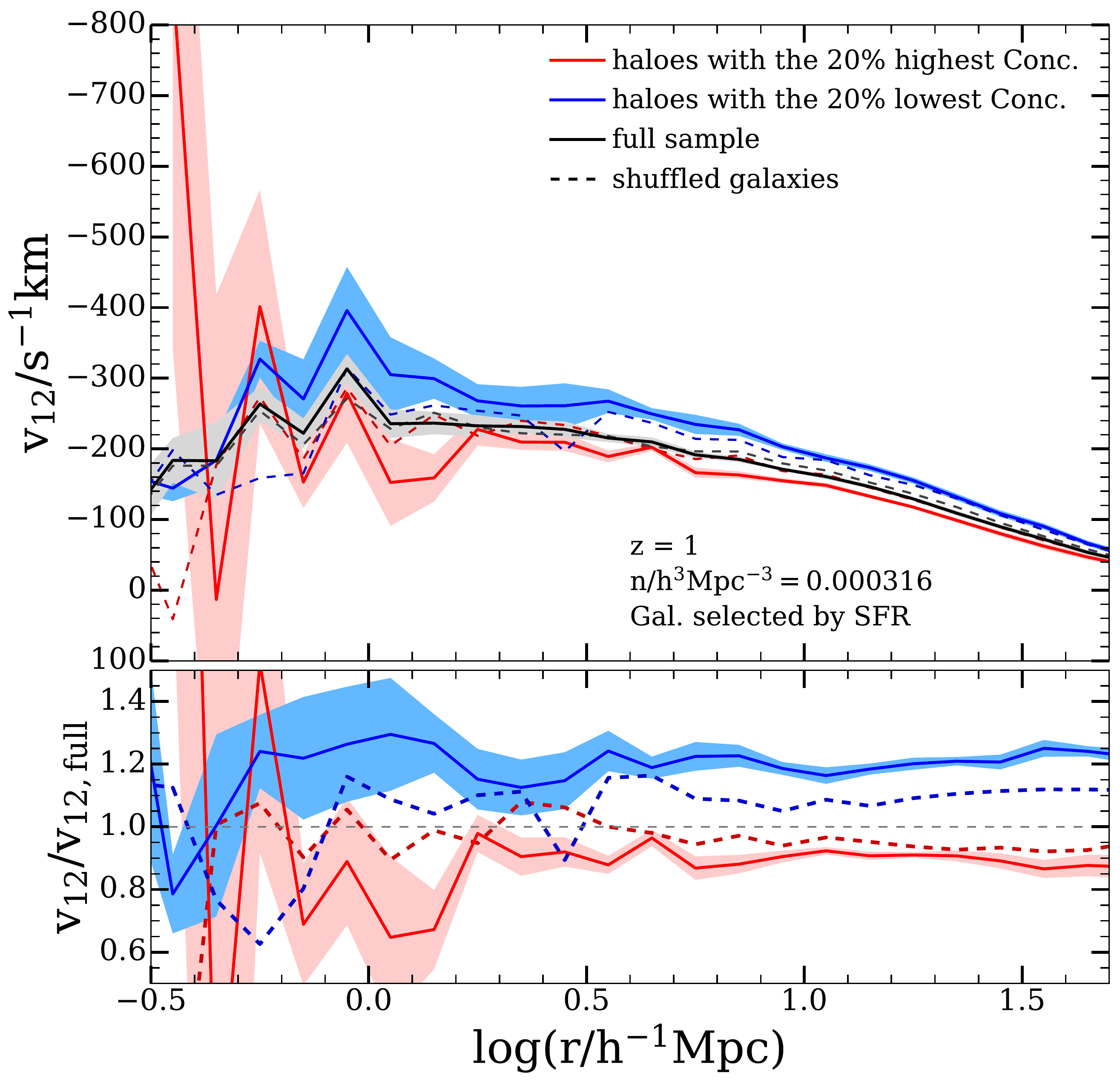}
\includegraphics[width=0.47\textwidth]{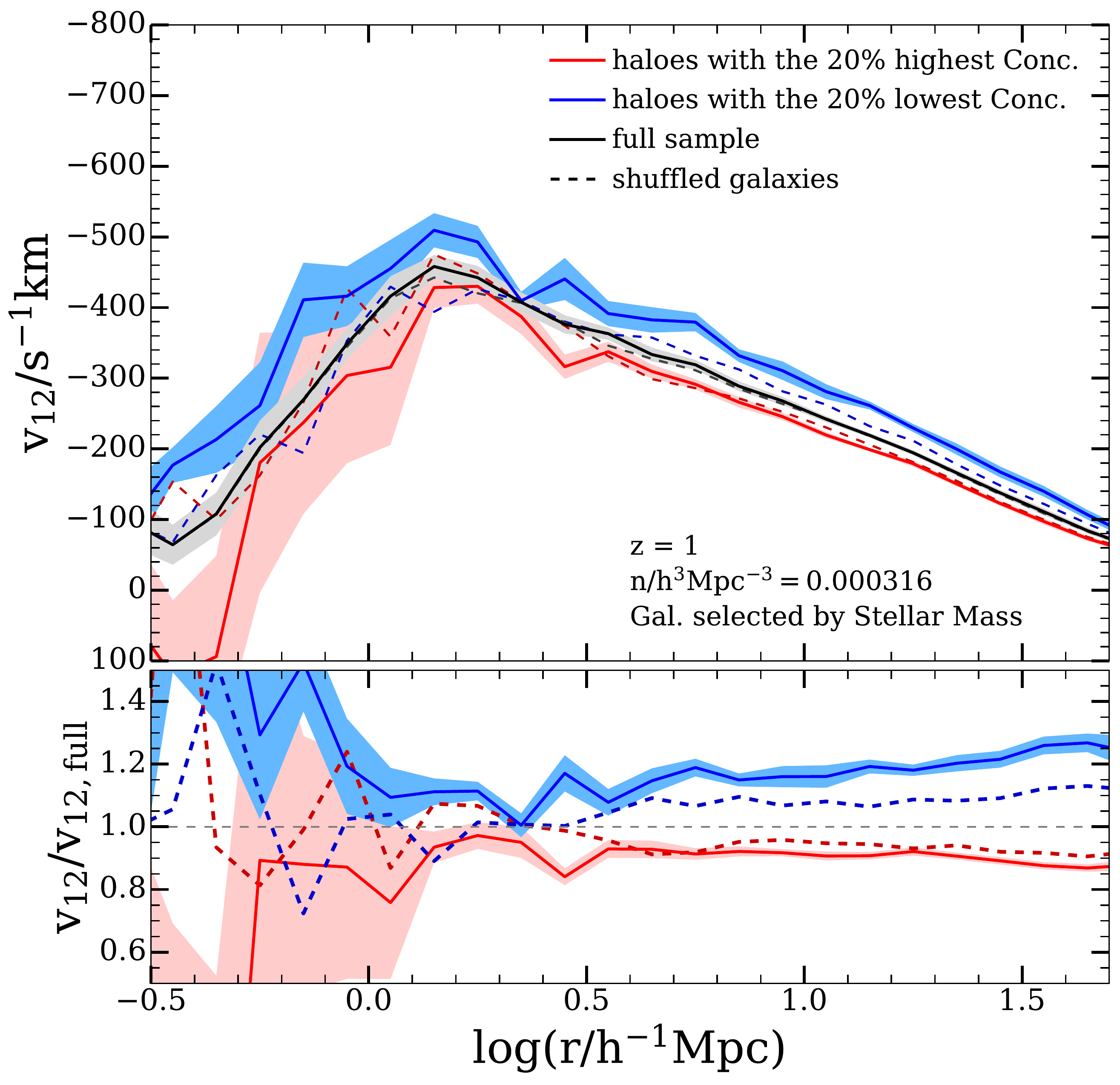}
\includegraphics[width=0.47\textwidth]{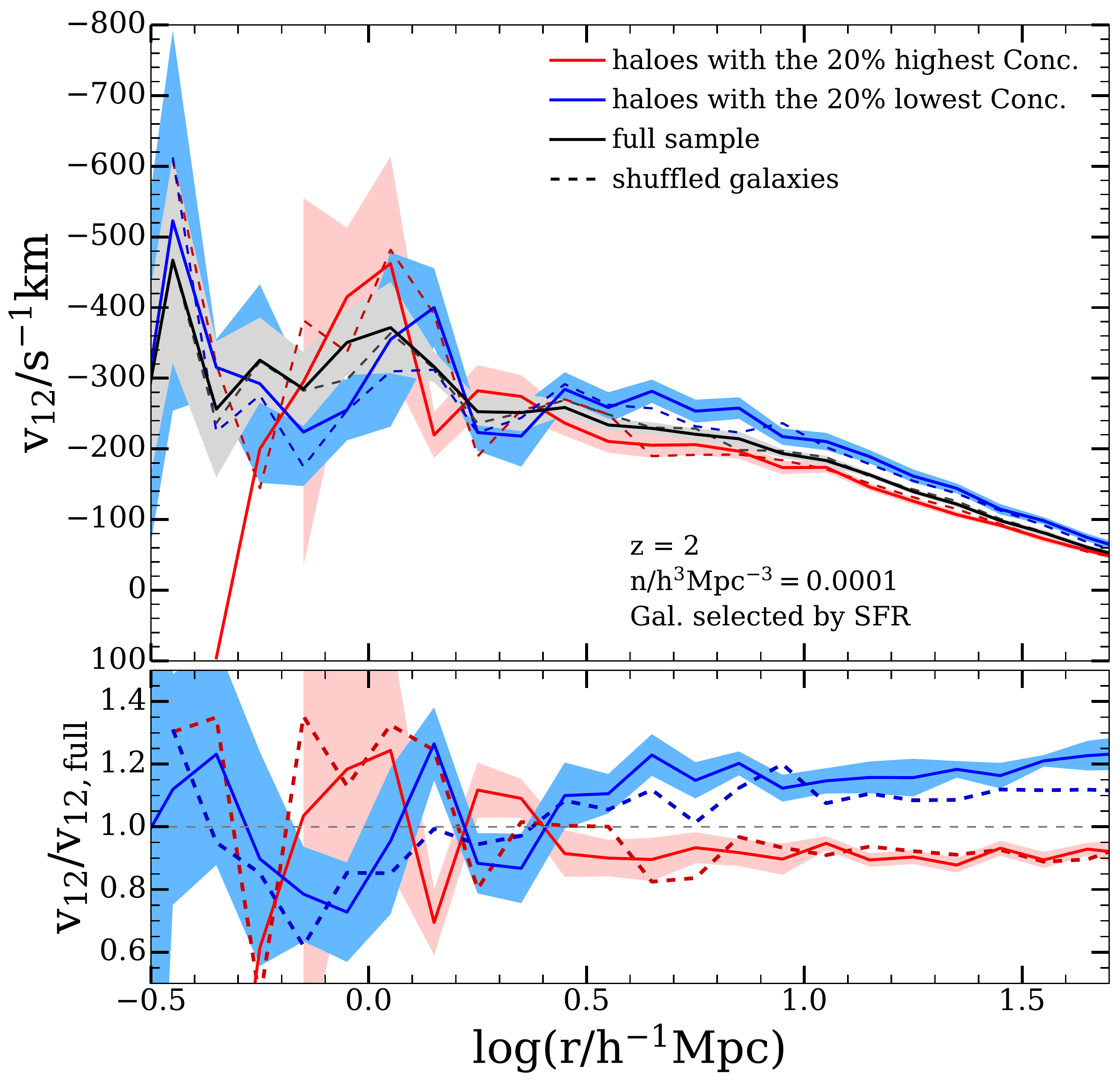}
\caption{Pairwise velocities for all galaxies in samples selected by stellar mass for $z=0$ and $z=1$ (top and bottom left panels, representing the SDSS main galaxies and a $z=1$ CMASS-like samples, respectively), and galaxies selected by star formation rate for $z=1$ and $z=2$ (top and bottom right panels, aimed at representing the SDSS-IV ELG and HETDEX samples, respectively), with space densities indicated in the key.  The solid lines show the results for SAM galaxies in their original haloes, whereas the dashed lines are for the catalogues where galaxies where shuffled among haloes retaining the halo spatial and velocity structures.  Black lines
are for the full samples, red and blue for high and low concentrations, respectively.}
\label{Fig:vcstm}
\end{figure*}

Figure \ref{Fig:vcstm} shows the pairwise velocities for selected samples both in their original and shuffled forms (in solid and dashed lines).  Negative values correspond to pairs of galaxies that are moving towards one another. The case where pairwise velocities are calculated between all galaxies is shown in black; blue and red are for the $20\%$ 
lowest and $20\%$ highest concentrations against the entire sample,
respectively.  Solid lines show the original samples, dashed represent the shuffled catalogues.
The shaded regions show the error on the mean obtained from $10$ jackknife resamplings. 

The bottom subdivisions in each panel show the ratio of the high and low concentration samples to the full ones, with the same line style as the main panels.  The ratios are intended to highlight whether halo assembly bias or occupancy variation are responsible for
the difference in pairwise velocity amplitude.  If at large separations the ratio of shuffled samples
is significantly different than unity, then halo assembly bias is behind the amplitude differences.  Only when
occupancy variation is  solely responsible will this ratio be equal to one.

The panels of the figure correspond to selections intended to represent the SDSS Main Galaxy sample
(top left), the Emission Line Galaxies of eBOSS (top right), a $z=1$ CMASS-like sample (bottom left)
and a HETDEX like survey (bottom right); the space densities and selection criteria are shown
by the key in each panel.
The bottom-left panel and the top-right panel correspond to the same redshift and space density but for different selections, stellar mass and star formation rate, and highlights the importance of
taking into account the physical property on which the selection is performed.

The pairwise velocity function shows a similar shape for the different samples, at least qualitatively.
At small separations it starts off with a low amplitude infall that increases toward the transition to 2-halo
term scales.  This is mostly due to the mix of first infall satellites with the ones that are in virial 
equilibrium within the halo; the fraction of first infalls increases with separation.  At larger
separations, $r \sim 3\ {h}^{-1}$Mpc and above, infall velocities drop monotonically as the enclosed overdensities 
gradually become smaller.  These results are qualitatively similar to the pairwise velocities of dark matter haloes reported by \cite{Xu:2018}.

In all cases, the galaxies residing in haloes in the lowest concentration quintile show the higher velocity infalls on the two-halo term, regardless of whether the selection is done using stellar mass or star formation rate.  The amplitude of the difference is similar regardless of redshift, space density, or selection by either stellar mass or star formation rate.  This trend extends down to the 1-halo term, although it is noisier for the star formation rate selections, and for the lower density samples.
However, since the selection is made using different galaxy properties, the average velocity values are different.  In particular, the stellar mass selection shows a higher amplitude of infall.

At equal space density and redshift (bottom left and top right panels of Fig. \ref{Fig:vcstm}) 
the amplitude of infall in the 1-halo term is larger 
for the star formation rate selection, but the difference between low and high 
concentration is less significant than in the stellar mass selection case.  This 
 points to a larger spread in infall amplitudes when star forming galaxies are 
selected.  We do not pursue the reasons for this difference in this present study.

The different panels of Figure \ref{Fig:vcstm} show the effect of assembly bias and occupancy variation in the different samples.
As can be seen, 
the $z=0$ case of a stellar mass selected sample 
illustrating the SDSS Legacy Sample, shows pairwise velocities 
that differ by $\sim 20\%$  on 2-halo term scales for high and low concentrations,
but almost no difference for the shuffled catalogues separated by 
concentration.  This indicates that an important fraction of the effect comes from occupancy
variation rather than halo assembly bias on its own.  
However, in the other panels, the shuffled and original catalogues present 
more similar behaviours, and these become almost the same for the $z=2$ case.  
This points to a trend of decreasing importance of occupancy variation as the 
redshift increases.  This is in agreement with \cite{C18} who report a similar trend 
of a smaller effect on the correlation function from 
occupancy variation as the redshift increases,
 for samples with low and high concentration, for both 
SFR and stellar mass selected samples.

For samples selected by halo age, for the same cases shown in this figure, the results are qualitatively similar.  For the SFR
selected samples, irrespective of redshift, the impact on the 2-halo term is smaller than for the concentration cuts for both halo assembly bias and occupancy variation; these results are shown in Figure \ref{Fig:vcstm2} of the Appendix.

\begin{figure}
\includegraphics[width=0.47\textwidth]{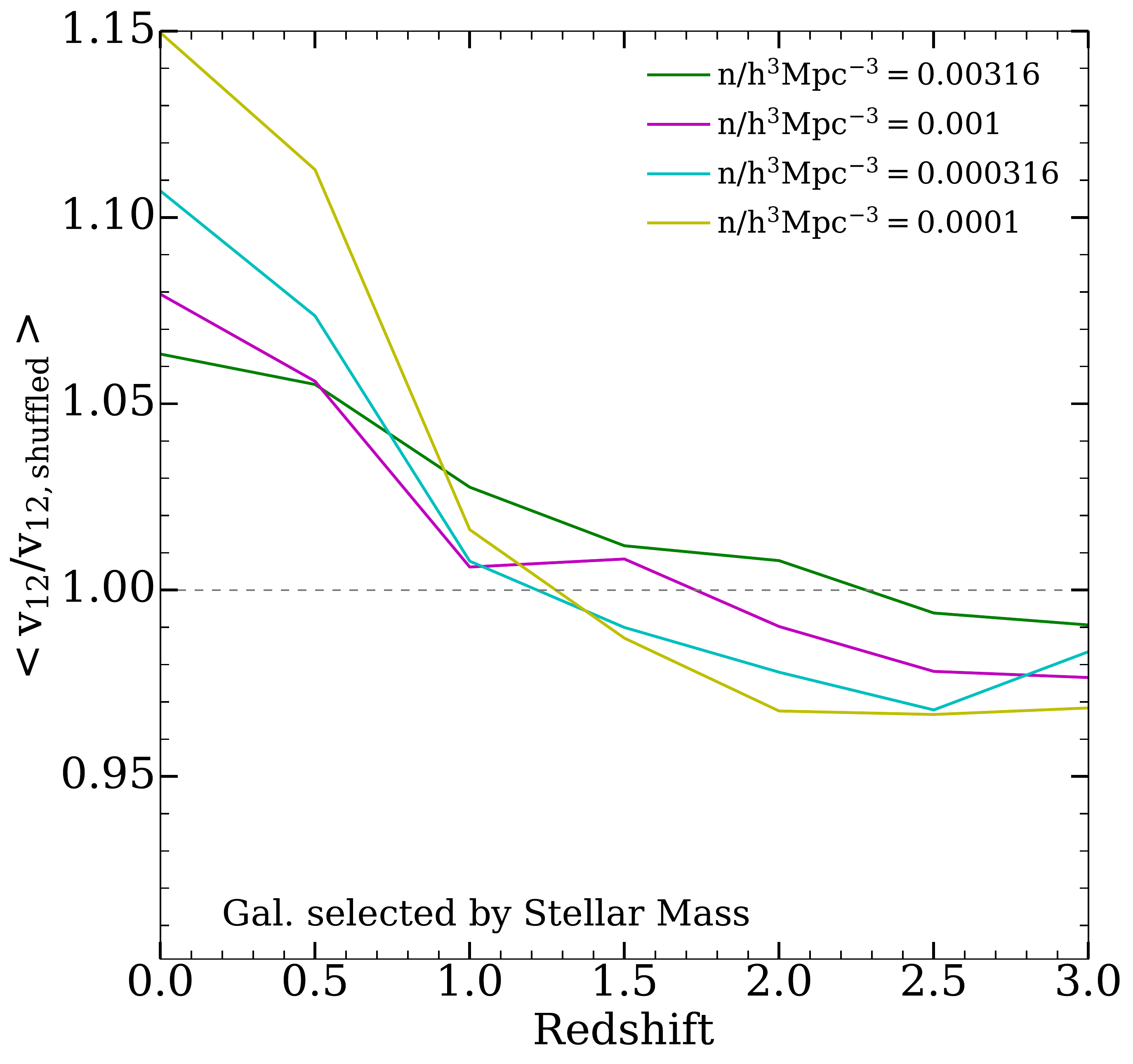}
\includegraphics[width=0.47\textwidth]{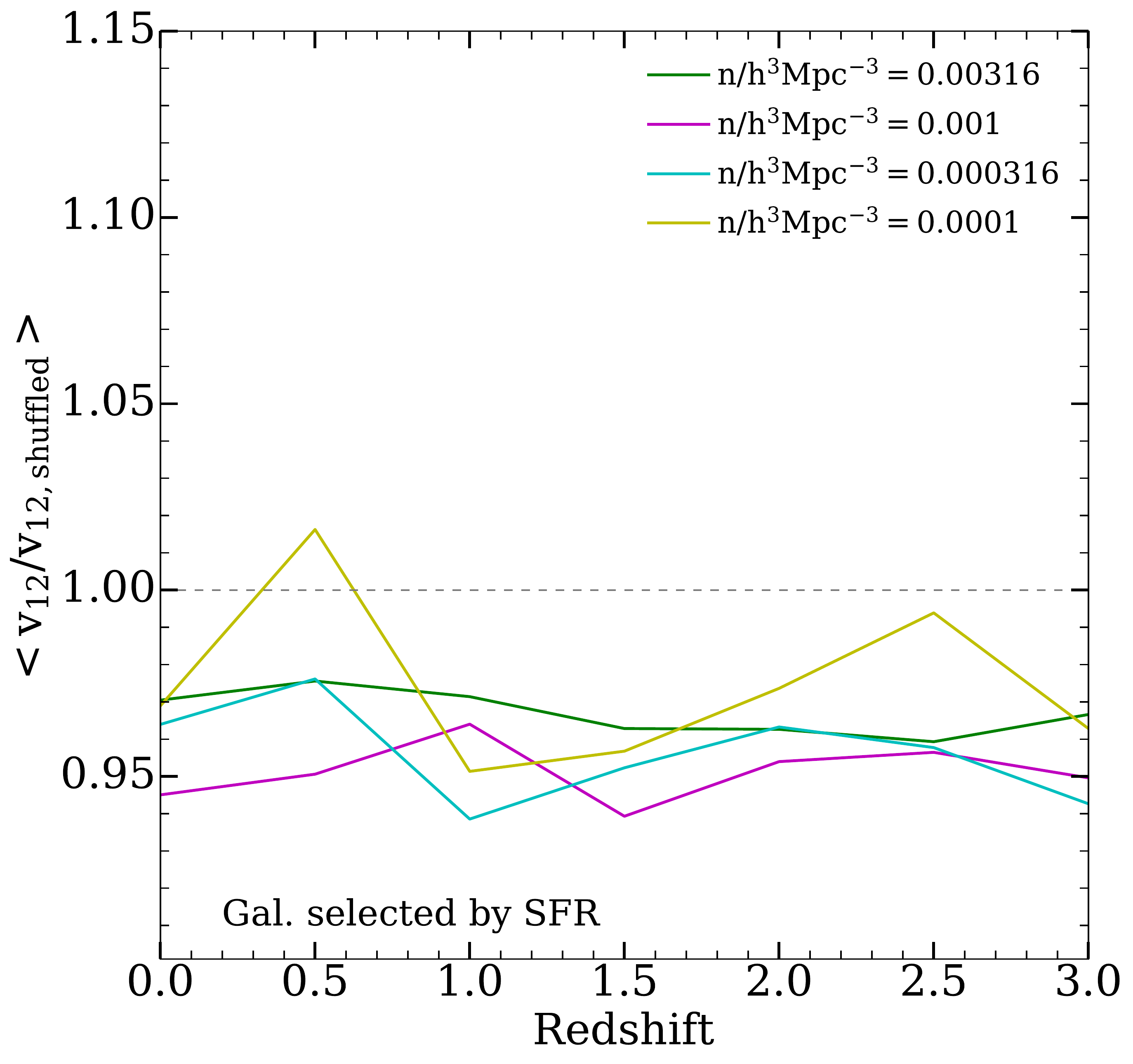}
\caption{
Ratio between averaged pairwise velocities over $5<r/\ {h}^{-1}$Mpc$<35$ of the original and shuffled 
catalogues for stellar mass and star formation rate selections (top and bottom panels) for different galaxy number
densities (different colours, indicated in the key). 
}
\label{Fig:vov}
\end{figure}
\begin{figure}
\includegraphics[width=0.45\textwidth]{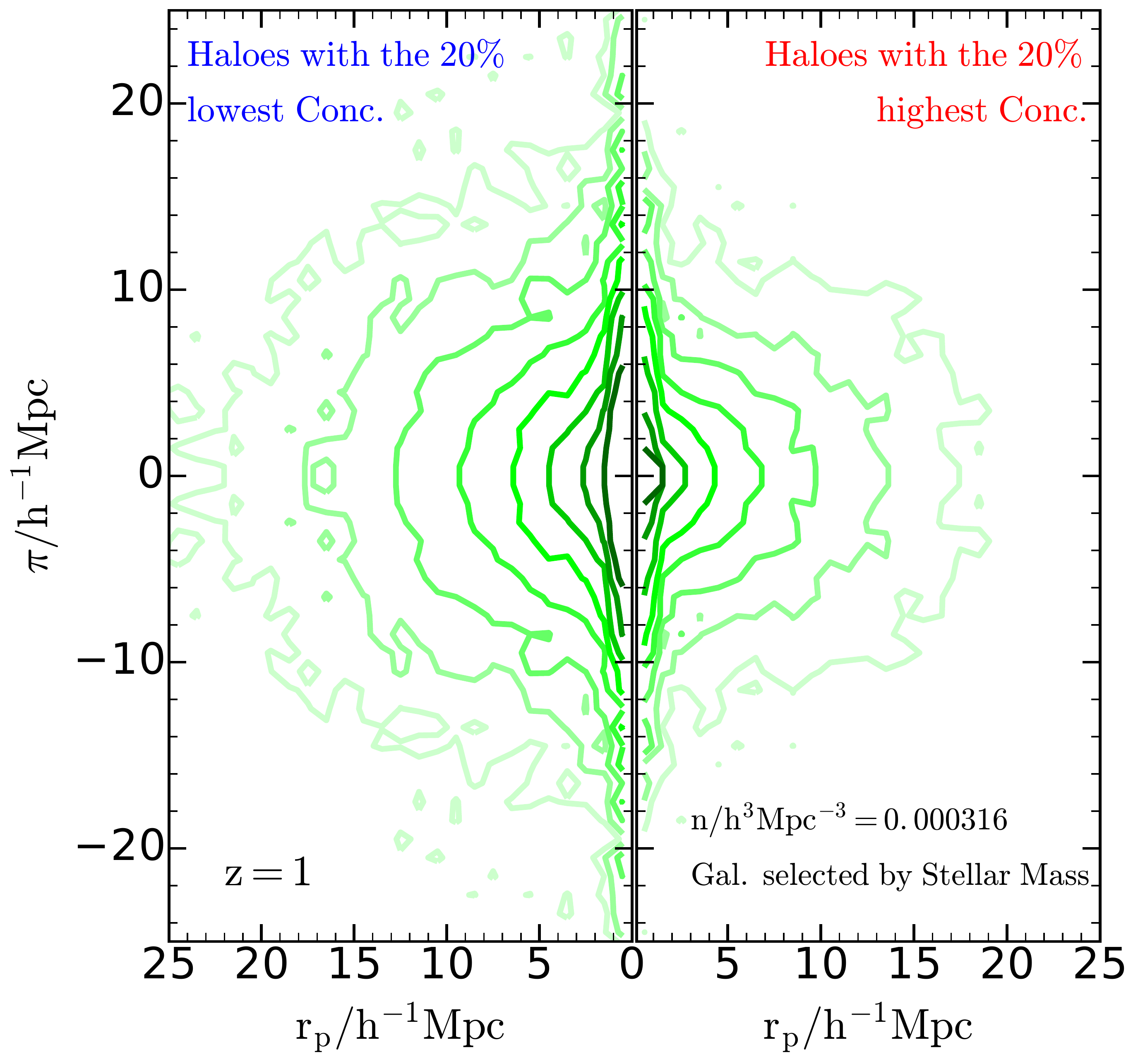}
\includegraphics[width=0.45\textwidth]{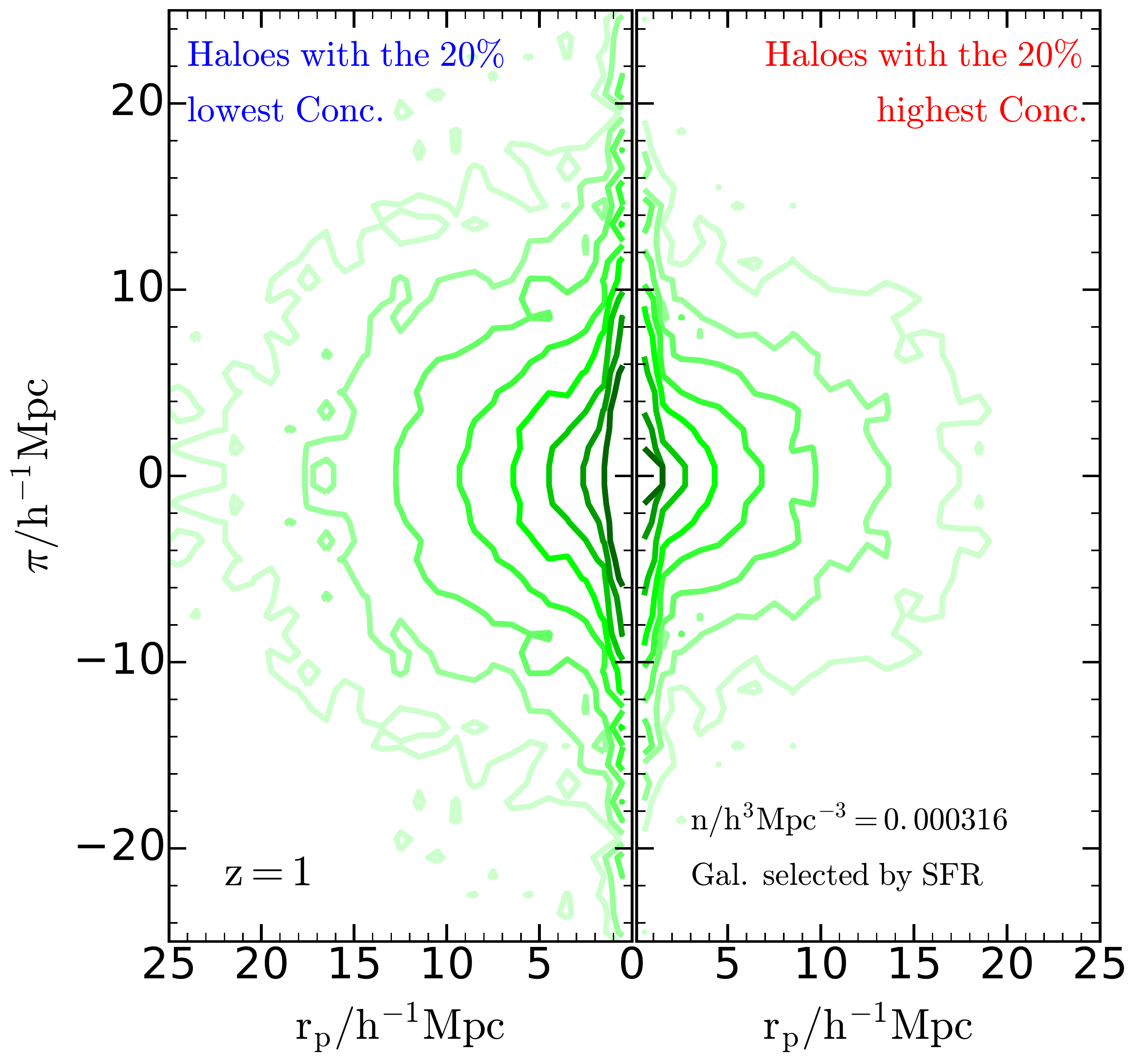}
\caption{Correlation functions as a function of separation in the directions parallel ($\pi$) and perpendicular to the line of sight ($r_p$), in this case coincident with the $z$ axis of the simulation box. Both panels correspond to $z=1$ and a space density of $n=0.000316$ ${h}^3$Mpc$^{-3}$, for stellar mass
and star formation rate selected samples (top and bottom, respectively), for low and high
concentrations (left and right, respectively).  The lines show contours of
equal correlation function amplitudes; darker shades are for higher amplitudes of $\log\xi=-0.5,-0.25,0,0.25,0.5,0.75,1$ and $1.25$.}
\label{Fig:xisigpi}
\end{figure}

Also in the Appendix, we redo this analysis using just the central galaxies, and find only small differences with respect to the ones presented in Fig. \ref{Fig:vcstm}, of the order of $10\%$.  The results for the stellar mass and star formation rate samples at $z=1$ and $n=0.000316\ {h}^3$Mpc$^{-3}$ can be seen in Fig. \ref{Fig:vccen}. 
This is consistent with \cite{Zehavi:2018} who also find that the assembly bias in clustering is dominated by the central galaxy population.

\subsection{Assembly bias and pairwise velocities}

We also study the evolution of the effect of assembly bias on pairwise velocities.
Figure \ref{Fig:vov} shows the ratio of the average values of the
pairwise velocities for the original and shuffled
catalogues as a function of redshift; note that here we are not dividing the sample into high and low values of the secondary parameter.
When comparing between original and shuffled results we are effectively looking at the fraction of the effect coming from occupancy variation since the shuffled samples are not affected by occupancy variation. 
The averaging is done on the 2-halo term range of separations 
$5<r/{h}^{-1}$Mpc$<35$.   By construction of the shuffled catalogues, this ratio is 1 in the 1-halo term regime.
The top panel shows the results for stellar mass selected samples of different space densities, whereas
the bottom panel shows this for star formation rate selections.

At low redshift, $z<1$, for the stellar mass selection (top panel), the infall is stronger
for the original catalogues irrespective of space density.  This indicates that the previously reported
effect of assembly bias that results in a stronger clustering
for the unshuffled samples (\citealt{C18}, \citealt{Zehavi:2018}) is also present for the relative velocities, which
show a stronger infall prior to shuffling consistent with larger mass overdensities.
As we move to higher redshifts we see that the importance of the occupancy variation becomes negligible, but as we reach $z>2$, the infall becomes
less pronounced in the unshuffled samples, a reversal of the behaviour at low redshift.  This is also
consistent with the reversal of the effect of occupancy variation seen in the clustering amplitude
by \cite{C18}.

The bottom panel shows the same calculation for SFR selected samples.  As can be seen, 
there is no clear dependence on number density or redshift, and there is a 
tendency to exhibit weaker large-scale infall in the original samples.  
This points in the same direction as the results of \cite{C18}
where the shuffled samples show more clustering than the unshuffled ones for SFR selected samples.

The fact that the changes due to assembly bias in the infall velocities are consistent with those
reported for the clustering suggests that the reason behind both is the change in the bias factor of the samples, as also
pointed out by \cite{Xu:2018}.
Later on in this work we will quantify whether the velocity differences can be explained by the change in the amplitude of clustering alone, i.e., by the change in the bias factor.  

\begin{figure}
\includegraphics[width=0.48\textwidth]{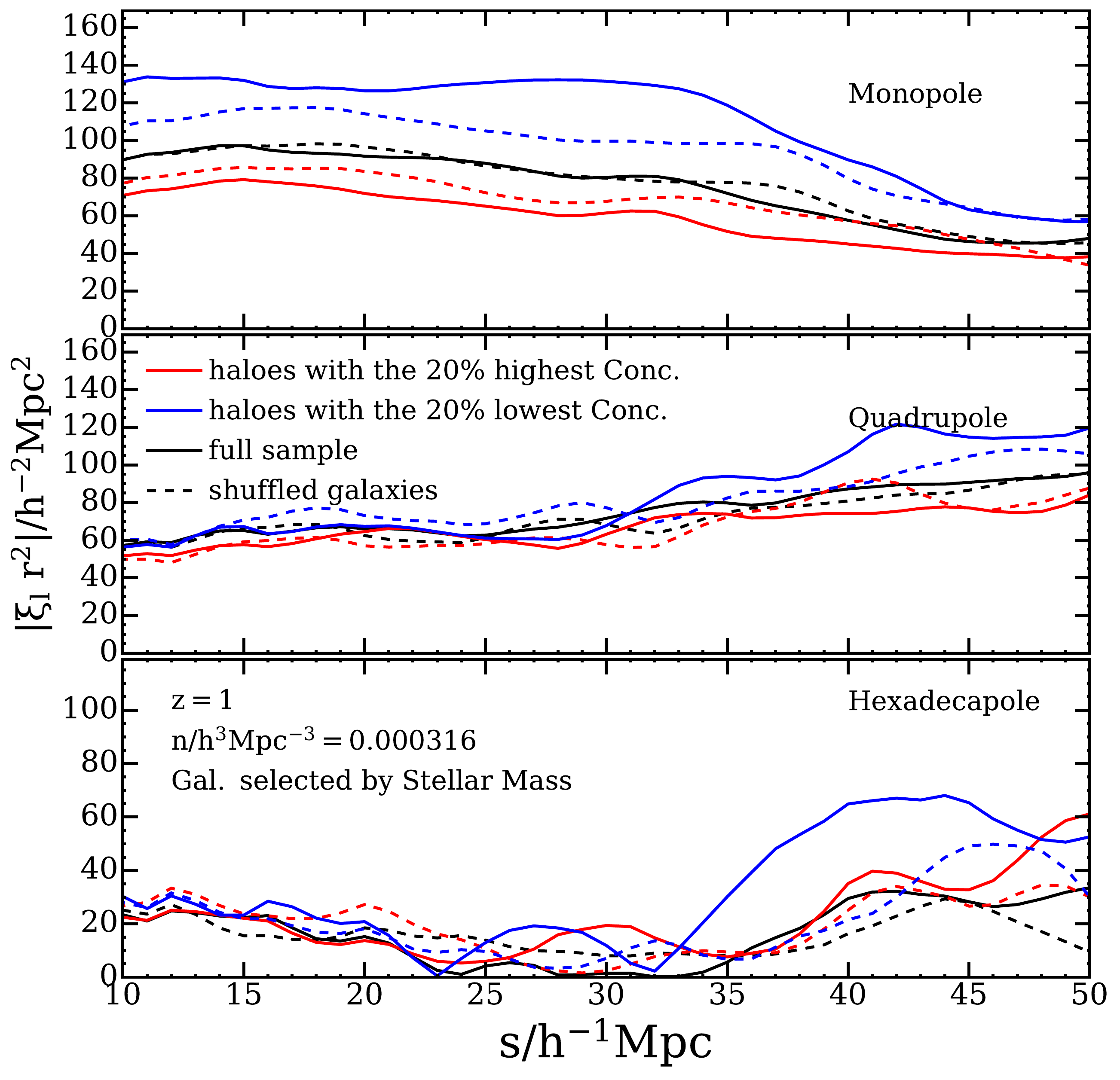}
\includegraphics[width=0.48\textwidth]{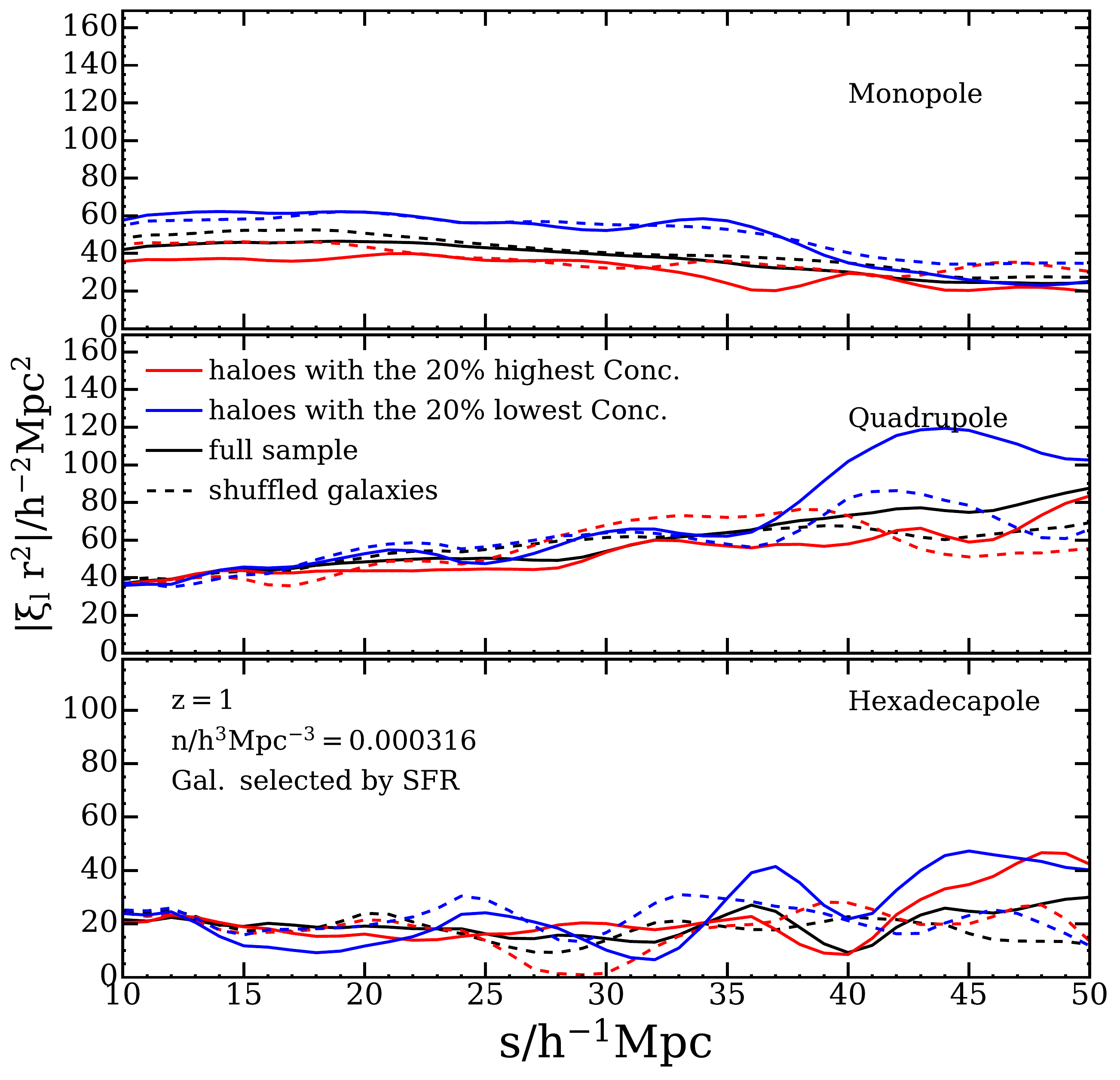}
\caption{Multipoles of the cross correlation function for $z=1$ and a space density of 
$n=0.000316\ {h}^3$Mpc$^{-3}$, for stellar mass (top) and star formation rate (bottom) selected
samples.  Each panel is further subdivided to show the monopole, quadrupole and hexadecapole ($l=0,1,2$) as a function of separation in redshift space $s$.  
Black lines show the full sample, blue and red the first and fifth quintiles in concentration,
and solid and dashed lines the original and shuffled samples, respectively.
}
\label{Fig:Multipoles}
\end{figure}

\section{Analysis in Redshift-space}
\label{Sec:Mult}

Before delving into the quantitative analysis of the multipoles of the correlation function
in redshift space, we will make a qualitative assessment of redshift space distortions for
samples of equal halo mass but different secondary properties.  

\begin{figure}
\includegraphics[width=0.44\textwidth]{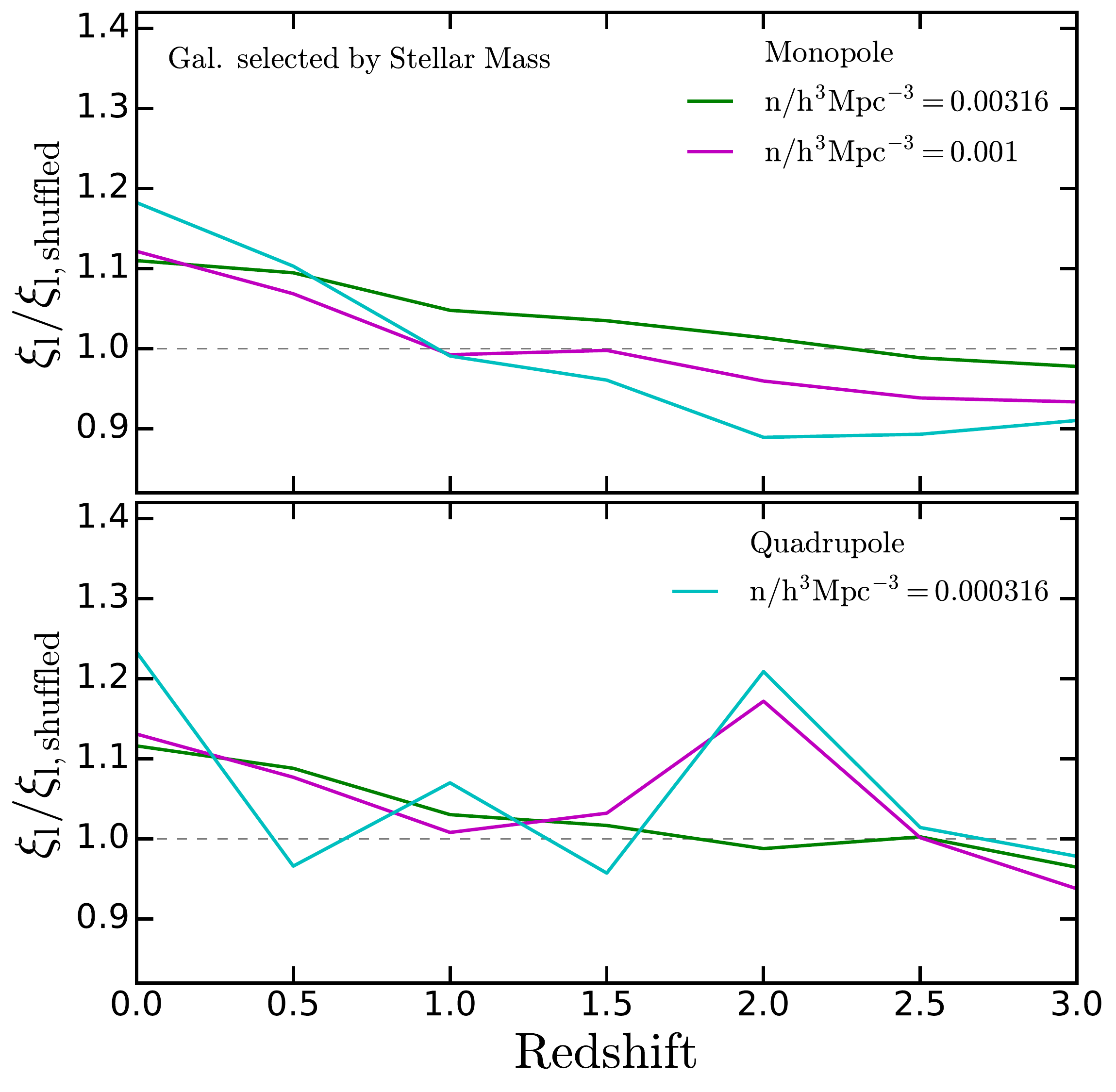}
\includegraphics[width=0.44\textwidth]{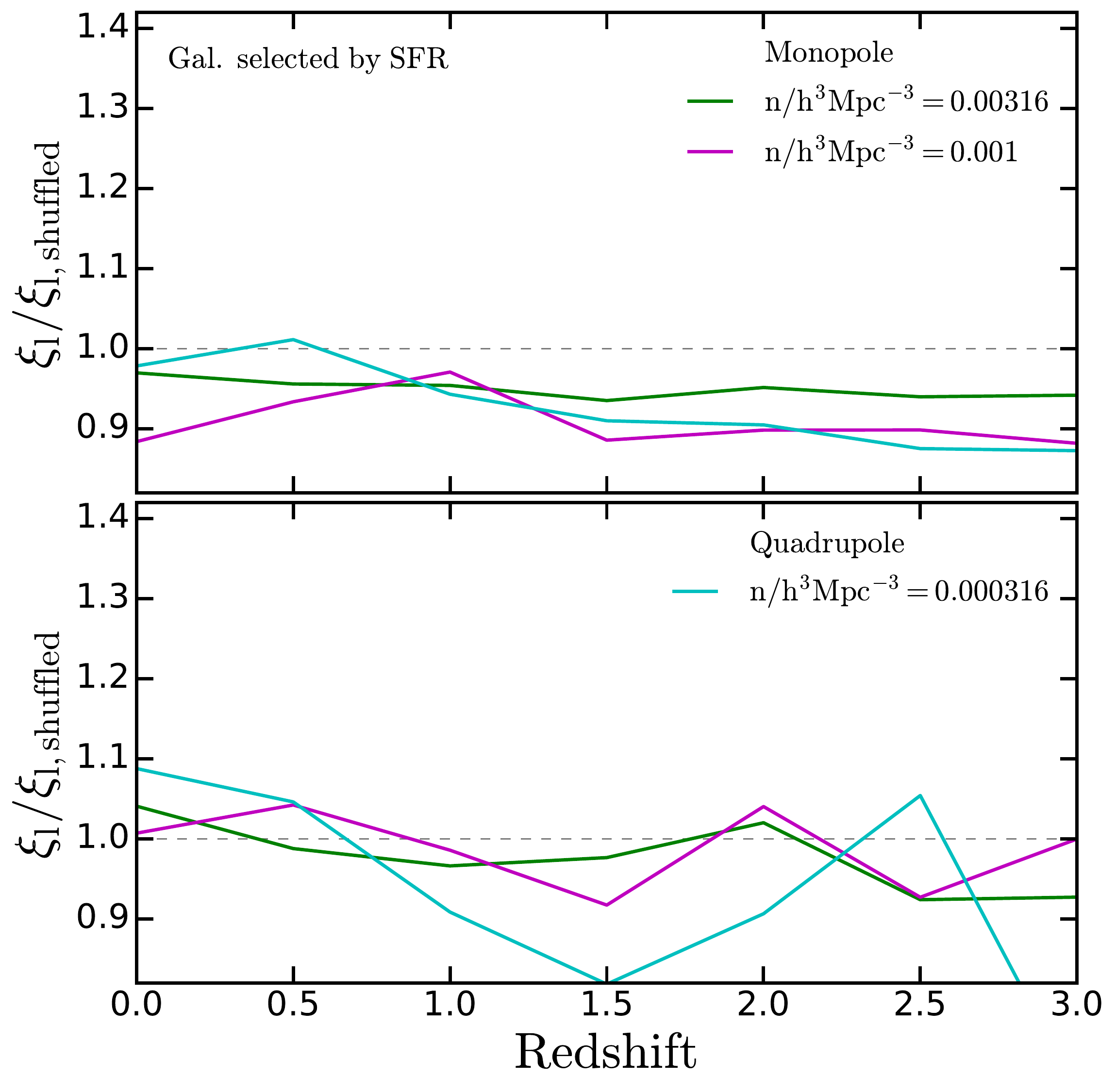}
\caption{ 
Ratio between averaged multipole moments of the redshift-space correlation functions for $5<r/{h}^{-1}$Mpc$<35$; this range isolates the scales on which the 2-halo term is important.
The ratios are shown as a function of redshift for different space densities (colours as indicated
in the key) and for stellar mass (top) and star formation rate selections (bottom).  The top and bottom subdivisions
of panels correspond to the monopole and quadrupole, respectively.
}
\label{Fig:multratio}
\end{figure}

\subsection{Redshift-space distortions of the correlation function}
We measure the correlation
functions as a function of projected and line of sight separations, $\xi(r_p,\pi)$ using the
 positions of galaxies in the simulation with the $z-axis$ position distorted by the peculiar
 velocities as mentioned above.  We calculate  auto-correlations for
our samples of different space densities and redshifts, and the corresponding 
cross-correlation functions between these full samples and quintiles with high or low concentration.  This allows us to see
the effect of the secondary halo parameter, i.e. assembly bias, on the redshift-space correlation functions.

Figure \ref{Fig:xisigpi} shows the actual redshift space distortion pattern for low and high halo concentrations (left and right panels, respectively) for stellar mass and star formation rate selected samples
(top and bottom) at $z=1$, with equal space densities.  
This figure shows the typical pattern of redshift space distortions, where at small
projected separations the equal correlation function amplitude contours are elongated in the direction
of the line-of-sight due to the random motions inside virialized structures (the so called 
``fingers-of-God" effect).  At larger projected
separations, the contours become squashed in the line-of-sight due to the coherent large-scale
infall motions.

As can be seen
the elongations at small $r_p$ are present in all cases, but there are clear differences in both the amplitude and redshift-space distortion of the correlation functions
between high and low concentration samples where the latter show a less prominent infall pattern (squashing
of contours).  There are also hints of a stronger infall pattern in the samples selected by the galaxy
star formation rate but this is probably due to a combination of smaller
fingers-of-God effect and the actual infall, because the amplitude of 
the correlation function is higher for the stellar mass selection, as is also
the case for the amplitude of the pairwise velocities (see Figure \ref{Fig:vcstm}) .
Left panels with the low concentration samples show a higher clustering amplitude.

\subsection{Multipoles}

We quantify the effect of redshift space distortions 
on the correlation function by measuring the multipoles $\xi_l(s)$,
\begin{equation}
\xi_l(s)=\frac{2l+1}2 \int {\rm d}\mu \xi(r_p,\pi) L_l(\mu),
\label{Eq:multipoles}
\end{equation}
where $s=\sqrt{r_p^2+\pi^2}$ is the separation in redshift space, $l$ is the 
multipole moment, $\mu=\pi/s$, and $L_l(\mu)$ are Legendre polynomials of order 
$l$ (\citealt{Hamilton:1992,Cole:1994,Padmanabhan:2008}).  
Figure \ref{Fig:Multipoles} shows the monopole, quadrupole and hexadecapole 
($l=0,2,4$, respectively) for the samples of galaxies selected by stellar mass and 
star formation rate (top and bottom panels) at $z=1$ for a space density of 
$n=0.000316\ {h}^3$Mpc$^{-3}$. Solid lines correspond to the
original samples, and dashed lines to the shuffled samples.

Irrespective of whether star formation rate or stellar mass is used to select the samples, we find
that the monopole shows results that are consistent with the real space correlations reported by \cite{C18} for both assembly bias and occupancy variation.
Namely that the low concentration samples show stronger clustering.  This 
was not necessarily  expected  since
the redshift-space monopole responds in some degree to redshift space distortions
which are also influenced by  assembly bias and  occupancy variation, as shown in the previous section.
This effect is stronger for the stellar mass selection.
The shuffled catalogues continue to show this trend but to a lesser degree; the small differences
with respect to the full samples show that there is an effect from assembly bias.  

The quadrupole is entirely due to the effect of redshift space distortions.  
For both stellar mass and SFR selections
the quadrupole shows a higher amplitude for the low concentration samples, particularly for the largest
scales $r/{h}^{-1}$Mpc$>30$.  The resulting quadrupoles for the shuffled samples appear qualitatively
similar but with a lower amplitude, pointing to an effect coming mostly from halo assembly bias as in the shuffled samples there is no occupancy variation.
The hexadecapoles are consistent between samples with low and high concentration, and for the shuffled and 
unshuffled catalogues, at least to within the uncertainties of our measurements.
By averaging the multipoles over 2-halo term scales these differences can be studied in more detail.

Fig.~\ref{Fig:multratio} shows the average values of the ratios between multipoles for the original and shuffled
results. We calculate the average over the range $5<r/{h}^{-1}$Mpc$<35$.  
The top panels show the ratio for stellar mass selected samples for three representative space densities.
As can be seen, the original samples show a stronger monopole at low redshifts and this difference decreases
for higher redshifts.  By $z=1$ the difference becomes less significant, and for the two lowest
space densities shown in the figure the difference tends to reverse making the monopole stronger for the shuffled samples.  The latter
suggests that occupancy variation tends to shift from preferentially weighting high mass haloes at low
redshift to low mass ones at $z>1$ thereby lowering the clustering amplitude.  This is similar to the behaviour
of the infall strength as shown in Figure \ref{Fig:vov}.  Furthermore, the ratio for the star formation rate selected
samples of the monopole for the original and shuffled samples are also consistent with the results for the
infall velocities.

On 2-halo term scales, the quadrupole is expected to show a direct relation with the infall velocities, as the latter
produce the flattening of the iso-correlation function contours that are quantified by the quadrupole.  As 
a result, we find similar trends to those shown in Figure \ref{Fig:vov} and to the results for the monopole.  Namely,
the effects from occupancy variation on clustering and infall velocities are consistent with one another, at least
qualitatively.
Even though the quadrupole showed stronger differences at
$r/{h}^{-1}$Mpc$>30$  (Figure \ref{Fig:Multipoles}), using this range of separations for the quadrupole does not influence our conclusions on the average ratios.

\section{Cosmological parameters: $\beta$}
\label{Sec:Cosm}

In the previous sections we showed the effects of assembly bias and occupancy variation on the pairwise
infall velocities and redshift space clustering of galaxies.  We found that in general there is
a qualitative agreement between the results for velocities and clustering in the sense that samples that
show stronger clustering also show stronger pairwise velocities as expected if bias is the main
driver of both effects.

In this section we make a quantitative comparison between the two by studying the resulting values
of the $\beta$ parameter.  This parameter represents the ratio between the logarithmic growth rate
$f(\Omega)$, which depends on the matter density parameter, and the bias parameter of a sample,
\begin{equation}
\beta=\frac{f(\Omega)}{b_s}.
\label{Eq:beta}
\end{equation}

\begin{figure}
\includegraphics[width=0.44\textwidth]{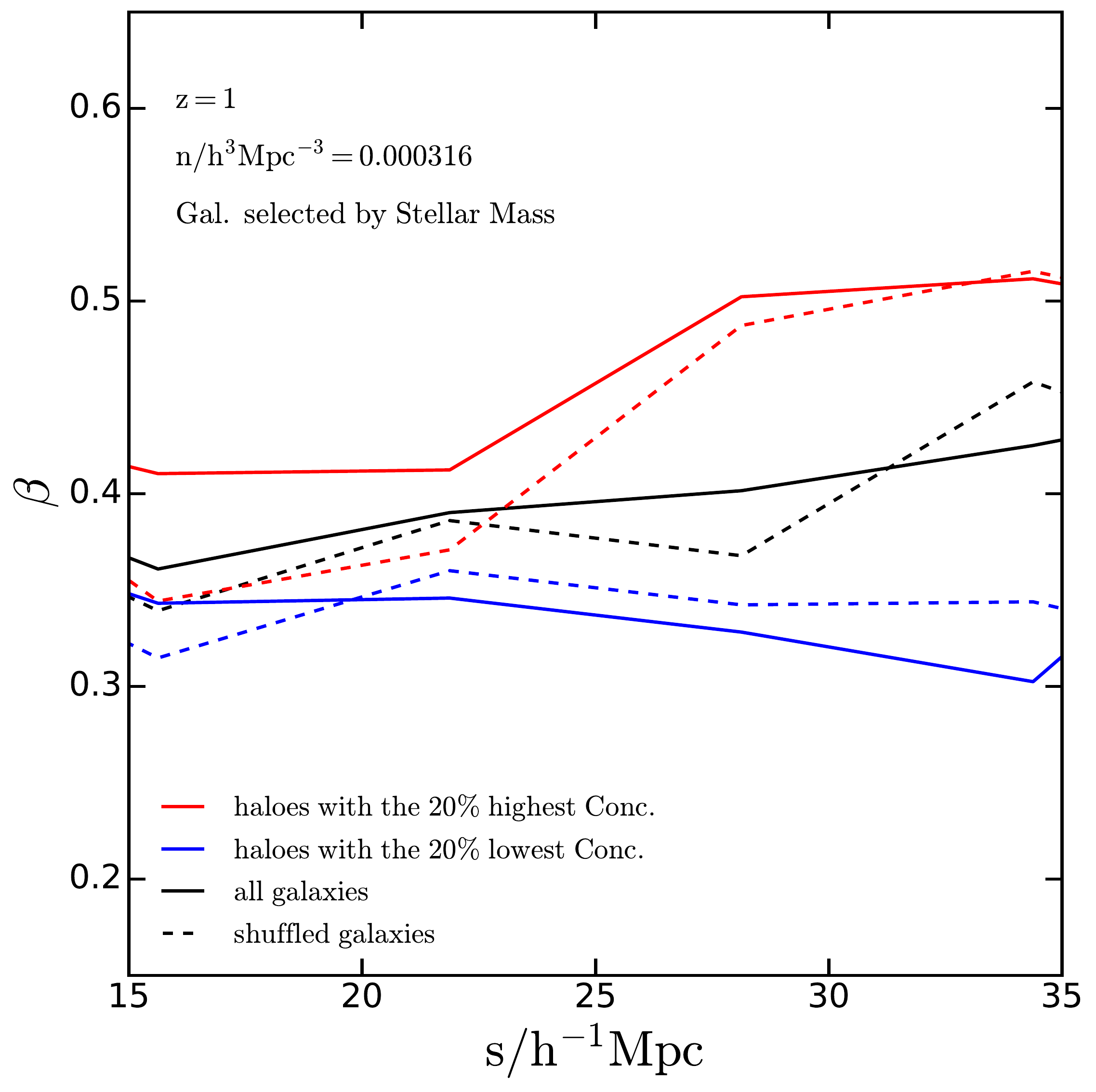}
\caption{Values of the $\beta$ parameter as a function of separation for a stellar mass selected sample with $n=0.000316\ {h}^3$Mpc$^{-3}$ (solid black line), and for the quintiles of high and
low halo concentration (solid red and blue lines, respectively), as a function of
separation.  The dashed lines are for the corresponding shuffled samples.}
\label{Fig:beta}
\end{figure}

The $\beta$ parameter can be obtained from the ratio between the monopoles of the correlation functions
in real and redshift space \citep{Kaiser:1986},
\begin{equation}
\xi_0(s)\equiv \left( 1+\frac23 \beta+\frac15 \beta^2 \right) \xi(r).
\end{equation}  
Since we are measuring cross-correlations, $b_s$ from Eq. \ref{Eq:beta} is the bias of the sample
with high or low halo concentration
\citep{Hamilton:1992,Peacock:1999}.
We show the values of this parameter in Fig.~\ref{Fig:beta} for a galaxy sample with 
$n=0.000316\ {h}^3$Mpc$^{-3}$ selected by their stellar mass at $z=1$ (illustrative
of a $z=1$ CMASS eBOSS-like galaxy sample).  We also measure $\beta$ using
multipoles as is done with real data; the results are shown in the Appendix (Figure \ref{Fig:betamult}).
We do not compare to the expected value of $\beta$ because rather than being interested in the
ability of this particular redshift-space distortion model to reproduce the measured value
of this parameter, we are interested in how its value changes due to halo assembly bias and
occupancy variation.  Different models for RSD will be tested in Jimenez et al. in preparation.

For this sample the ratio of 
the monopoles is lower for galaxies in low concentration haloes; note that this difference can be due
to the different bias factors of low and high concentration haloes.   
We also see that $\beta$ is roughly
constant between separations of $15$ and $35\ {h}^{-1}$Mpc as expected from linear theory. 
The dashed lines show the resulting values of $\beta$ for the shuffled catalogues (full, low and high concentration).  As can be seen the results for high and low concentration are similar for the
original and shuffled samples (little effect from occupancy variation).  The differences
between high and low concentration are consistent with the pairwise velocity results, and also
with \cite{C18}, giving an indication that the change in $\beta$ is driven by the change in the
bias factor of the sample.

\begin{figure}
\includegraphics[width=0.44\textwidth]{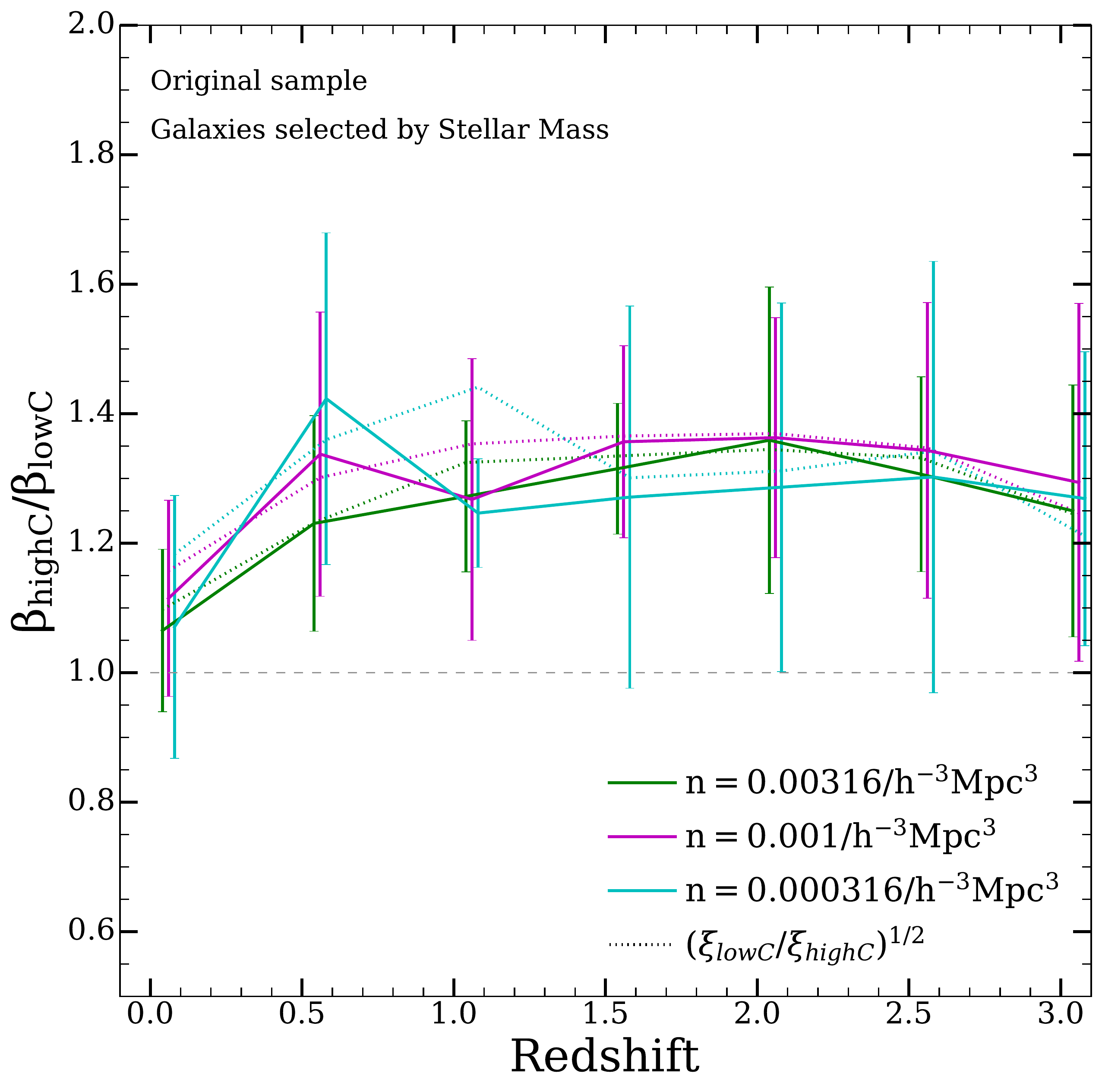}
\includegraphics[width=0.44\textwidth]{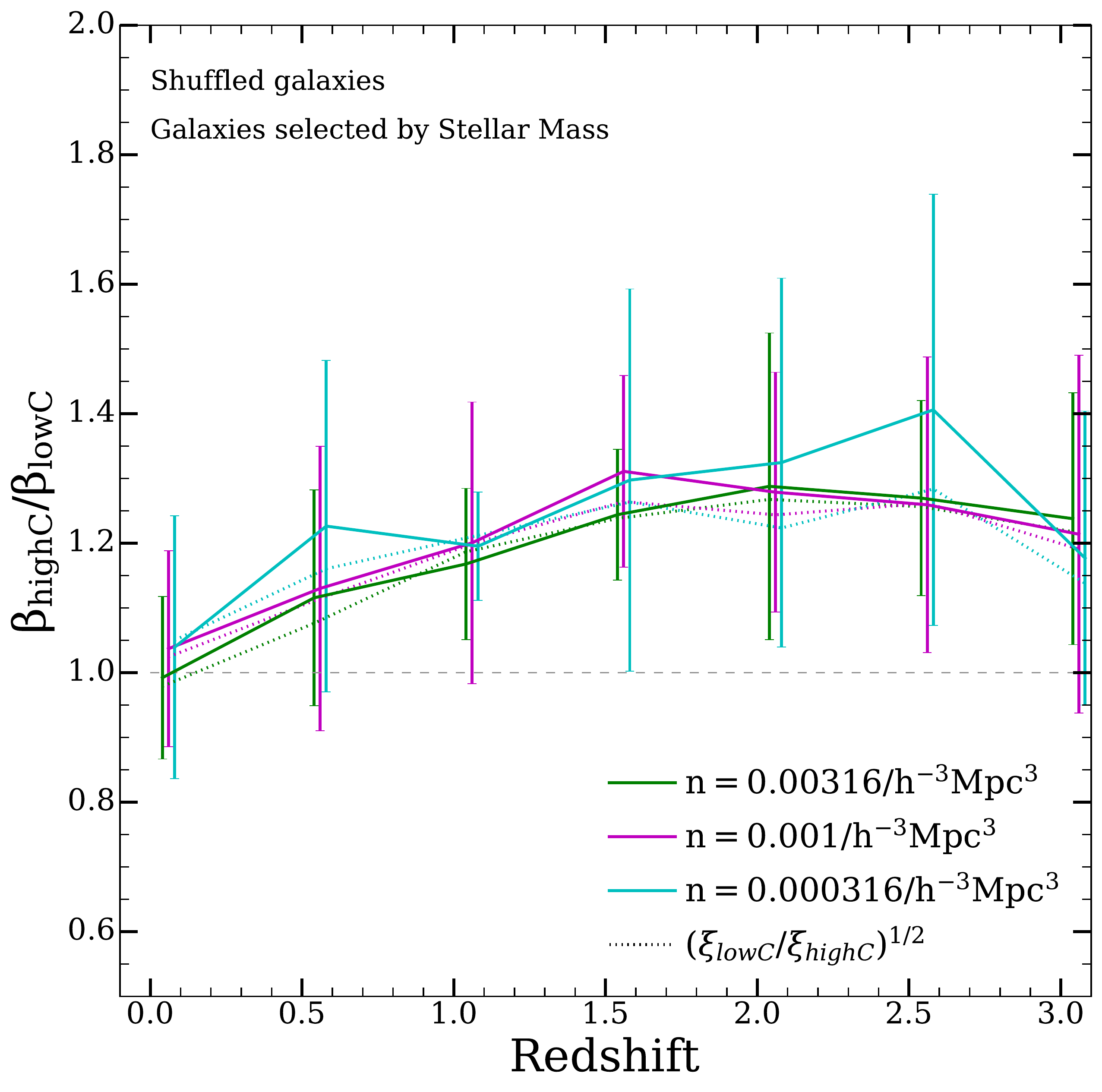}
\caption{Ratio between the averaged $\beta$ values for high and low concentration samples,
as a function of redshift, for different space density samples (colours, as indicated in the key)
selected by their stellar mass. Solid lines show results for the original samples; dashed lines
show the square root of the ratio between the real-space correlation functions for the same samples, averaged over the same range of separations; the agreement between solid and dashed lines show that both velocities and clustering respond to assembly bias in similar ways.  The horizontal grey dashed line shows the unit ratio.  The top panel correspond to the samples as directly obtained from the simulation, the bottom panel to the shuffled ones. Errors are obtained
from jackknife resamplings.}
\label{Fig:betaav}
\end{figure}

Figure \ref{Fig:betaav} shows the averaged ratio of $\beta$ for the high ($\beta_{\rm highC}$) and low concentration selection
($\beta_{\rm lowC}$), 
over the range $s=15-35\ {h}^{-1}$Mpc for different number density samples selected
by cuts in stellar mass (see the figure key) as a function of redshift.
As can be seen, this ratio evolves with redshift possibly due
to bias evolution as we evaluate below.
Its dependence is qualitatively similar
for the original and shuffled samples.  At $z=0$ the $\beta$ values for low and high concentrations 
are consistent with one another but by $z\sim 1$ the high concentration samples show a higher $\beta$ 
parameter by about a $20$ percent.  At $z=3$ in both the original and the shuffled samples, 
the ratio is again consistent with $\beta_{\rm highC}/\beta_{\rm lowC}=1$. Errorbars show the scatter in the ratios from jackknife resamplings.

The figure also shows as dashed lines the square root of the inverse 
ratio between the cross-correlation functions of the same samples, averaged
over the same range of separations.  This quantity corresponds to the ratio of bias factors
of the samples, i.e. the denominator of Eq. \ref{Eq:beta}.  As the dashed lines are consistent
with the solid ones, we can conclude that galaxy assembly bias does not introduce any detectable
systematics on estimates of $f(\Omega)$.
We have also repeated this analysis for SFR selected samples (see Figure \ref{Fig:betaavsfr} in the Appendix), reaching the same conclusion.

As galaxy assembly bias includes both the effect of halo assembly bias and occupancy variation, we also investigate whether halo assembly bias alone produces any systematic effects on $\beta$ in the bottom panel of Figure \ref{Fig:betaav}, which shows the result for the shuffled samples.  No systematic effects can be seen in these results either to the level of precision of our analysis (note
that this is lower than in \citealt{Chen:2018}).
This conclusion should also hold for more complex RSD modeling since according to these results
assembly bias and occupancy variation simply change the bias of the sample of galaxies, and it is possible to use
this value to obtain accurate estimates of total mass overdensities.  We will further investigate this in 
Jimenez et al. (in preparation).

\section{Conclusions}
\label{Sec:End}

We studied how galaxy assembly bias, the phenomenon by which the clustering of galaxies depends not
only on the halo occupation distribution and halo mass, but also on secondary halo properties, affects peculiar
velocities and their relation to the spatial distribution of galaxies.  We use the \cite{Guo:2013a} semi-analytic
model run on the Millenium WMAP-7 N-body simulation to select samples of galaxies using stellar mass and
star formation rate cuts, such that the space density of the samples spans several orders of
magnitude, with redshifts ranging from $z=0$ to $z=3$. We pay special attention to those that more closely mimic the SDSS main
galaxy sample, a CMASS-like and the ELG samples at $z=1$, and a $z=2$ HETDEX galaxies.  This work is the third in a series studying galaxy assembly bias in SAMs following \cite{Zehavi:2018} and \cite{C18}, where we move our focus from spatial clustering
to velocity space.

Our work complements \cite{Xu:2018}; while they concentrate on assembly bias of 
haloes, looking at several secondary properties, and how these produce correlated
changes in pairwise velocities and clustering of haloes, we focus on semi-analytic
galaxies, 
and study samples of equal halo mass and different concentrations.  We also test whether cosmological parameter
constraints using redshift space distortions could be affected by systematics when not taking into account the assembly
bias effect.

In general we find that the peculiar velocity field shows the expected response to halo assembly bias and occupancy variation; galaxy velocities respond to the bias factor of the sample regardless
of whether the halo mass or other halo property is driving the change in the bias
factor of the sample.  We summarize our results as follows:
\begin{itemize}
\item Low redshift, high space density, stellar mass selected samples such as the SDSS Main Galaxy
Sample, show strong occupancy variation effects on both the spatial correlations and pairwise velocities.  On the
other hand, halo assembly bias
gains importance for both stellar mass and SFR selected samples at $z=1$ such
as CMASS-like and ELGs.  At higher redshifts, the occupancy variation appears to have no effect on the
clustering and velocity variations, with most of the effect coming from halo assembly bias, at least for
the sparse sample that will be available at $z\sim 2$ from surveys such as HETDEX.  
\item The amplitude of the infall is affected by occupancy variation in a similar way as to what is 
seen in the  spatial clustering \citep{C18}; the 
shuffled catalogue shows a smaller difference of large
scale infall amplitude with concentration.
\item The results for the multipoles of the redshift-space correlation function show that both halo assembly
bias and occupancy variation affect the monopole and the quadrupole.  The size of the simulation does not allow us
to reach firm conclusions regarding the effect on the hexadecapole.  We use the monopole and the real-space correlation
function to estimate the $\beta$ parameter, and find that its changes are driven fully by
the variation of the bias parameter with halo mass or secondary halo property.  
\end{itemize}

To the level of accuracy allowed by the simulation, and to the level of the physics taken into account
by a semi-analytic model, we find that assembly bias does not introduce further systematics to cosmological
parameter estimates from redshift-space correlation functions in samples similar to the SDSS main galaxy sample,
$z=1$ CMASS-like or ELG eBOSS, or HETDEX samples.

Our results show that halo assembly bias, in particular, affects both clustering and velocities in a consistent way.  This is yet another indication that halo mass alone is  too simple a proxy for the equivalent
peak height of the original overdensities that evolved into haloes via complex
non-linear physics.

\section*{Acknowledgements}

We thank Raul Angulo and Alvaro Orsi for helpful comments and
discussions.
This work was made possible by the efforts of Gerard Lemson and
colleagues at the German Astronomical Virtual Observatory in setting
up the Millennium Simulation database in Garching.
NP \& SC acknowledge support from a STFC/Newton-CONICYT Fund award (ST/M007995/1 - DPI20140114) and Anillo ACT-1417.
NP is further supported by ``Centro de Astronom\'ia y Tecnolog\'ias Afines'' BASAL PFB-06
and by Fondecyt Regular 1150300. 
SC acknowledges support of the European Research Council through 
grant number ERC-StG/716151. 
IZ acknowledges support by NSF grant AST-1612085.
%
%
The calculations for this paper were performed on the 
ICC Cosmology Machine, which is part of the DiRAC-2 
Facility jointly funded by STFC, the Large Facilities 
Capital Fund of BIS, and Durham University and on the Geryon computer at the Center for 
Astro-Engineering UC, part of the BASAL PFB-06, which received additional
funding from QUIMAL 130008 and Fondequip AIC-57 for upgrades.

\bibliographystyle{mnras}
\bibliography{Biblio} 

\begin{thebibliography}{}
\makeatletter
\relax
\def\mn@urlcharsother{\let\do\@makeother \do\$\do\&\do\#\do\^\do\_\do\%\do\~}
\def\mn@doi{\begingroup\mn@urlcharsother \@ifnextchar [ {\mn@doi@}
  {\mn@doi@[]}}
\def\mn@doi@[#1]#2{\def\@tempa{#1}\ifx\@tempa\@empty \href
  {http://dx.doi.org/#2} {doi:#2}\else \href {http://dx.doi.org/#2} {#1}\fi
  \endgroup}
\def\mn@eprint#1#2{\mn@eprint@#1:#2::\@nil}
\def\mn@eprint@arXiv#1{\href {http://arxiv.org/abs/#1} {{\tt arXiv:#1}}}
\def\mn@eprint@dblp#1{\href {http://dblp.uni-trier.de/rec/bibtex/#1.xml}
  {dblp:#1}}
\def\mn@eprint@#1:#2:#3:#4\@nil{\def\@tempa {#1}\def\@tempb {#2}\def\@tempc
  {#3}\ifx \@tempc \@empty \let \@tempc \@tempb \let \@tempb \@tempa \fi \ifx
  \@tempb \@empty \def\@tempb {arXiv}\fi \@ifundefined
  {mn@eprint@\@tempb}{\@tempb:\@tempc}{\expandafter \expandafter \csname
  mn@eprint@\@tempb\endcsname \expandafter{\@tempc}}}

\bibitem[\protect\citeauthoryear{{Adams} et~al.,}{{Adams}
  et~al.}{2011}]{Adams:2011}
{Adams} J.~J.,  et~al., 2011, \mn@doi [\apjs] {10.1088/0067-0049/192/1/5},
  \href {http://adsabs.harvard.edu/abs/2011ApJS..192....5A} {192, 5}

\bibitem[\protect\citeauthoryear{{Artale}, {Zehavi}, {Contreras}  \&
  {Norberg}}{{Artale} et~al.}{2018}]{Artale:2018}
{Artale} M.~C.,  {Zehavi} I.,  {Contreras} S.,   {Norberg} P.,  2018, \mn@doi
  [\mnras] {10.1093/mnras/sty2110}, \href
  {http://adsabs.harvard.edu/abs/2018MNRAS.480.3978A} {480, 3978}

\bibitem[\protect\citeauthoryear{{Biviano} et~al.,}{{Biviano}
  et~al.}{2017}]{Biviano:2017}
{Biviano} A.,  et~al., 2017, \mn@doi [\aap] {10.1051/0004-6361/201731289},
  \href {http://adsabs.harvard.edu/abs/2017A%26A...607A..81B} {607, A81}

\bibitem[\protect\citeauthoryear{{Bullock}, {Dekel}, {Kolatt}, {Kravtsov},
  {Klypin}, {Porciani}  \& {Primack}}{{Bullock} et~al.}{2001}]{Bullock:2001}
{Bullock} J.~S.,  {Dekel} A.,  {Kolatt} T.~S.,  {Kravtsov} A.~V.,  {Klypin}
  A.~A.,  {Porciani} C.,   {Primack} J.~R.,  2001, \mn@doi [\apj]
  {10.1086/321477}, \href {http://adsabs.harvard.edu/abs/2001ApJ...555..240B}
  {555, 240}

\bibitem[\protect\citeauthoryear{{Busch} \& {White}}{{Busch} \&
  {White}}{2017}]{Busch:2017}
{Busch} P.,  {White} S.~D.~M.,  2017, preprint, \href
  {http://adsabs.harvard.edu/abs/2017arXiv170201682B} {} (\mn@eprint {arXiv}
  {1702.01682})

\bibitem[\protect\citeauthoryear{{Campbell}, {van den Bosch}, {Hearin},
  {Padmanabhan}, {Berlind}, {Mo}, {Tinker}  \& {Yang}}{{Campbell}
  et~al.}{2015}]{Campbell:2015b}
{Campbell} D.,  {van den Bosch} F.~C.,  {Hearin} A.,  {Padmanabhan} N.,
  {Berlind} A.,  {Mo} H.~J.,  {Tinker} J.,   {Yang} X.,  2015, \mn@doi [\mnras]
  {10.1093/mnras/stv1091}, \href
  {http://adsabs.harvard.edu/abs/2015MNRAS.452..444C} {452, 444}

\bibitem[\protect\citeauthoryear{{Chen}, {Zhang}, {Zheng}, {Yu}  \&
  {Jing}}{{Chen} et~al.}{2018}]{Chen:2018}
{Chen} J.,  {Zhang} P.,  {Zheng} Y.,  {Yu} Y.,   {Jing} Y.,  2018, \mn@doi
  [\apj] {10.3847/1538-4357/aaca2f}, \href
  {http://adsabs.harvard.edu/abs/2018ApJ...861...58C} {861, 58}

\bibitem[\protect\citeauthoryear{{Chuang} \& {Wang}}{{Chuang} \&
  {Wang}}{2012}]{Chuang:2012}
{Chuang} C.-H.,  {Wang} Y.,  2012, \mn@doi [\mnras]
  {10.1111/j.1365-2966.2012.21565.x}, \href
  {http://adsabs.harvard.edu/abs/2012MNRAS.426..226C} {426, 226}

\bibitem[\protect\citeauthoryear{{Cole}, {Aragon-Salamanca}, {Frenk}, {Navarro}
   \& {Zepf}}{{Cole} et~al.}{1994}]{Cole:1994}
{Cole} S.,  {Aragon-Salamanca} A.,  {Frenk} C.~S.,  {Navarro} J.~F.,   {Zepf}
  S.~E.,  1994, \mn@doi [\mnras] {10.1093/mnras/271.4.781}, \href
  {http://adsabs.harvard.edu/abs/1994MNRAS.271..781C} {271, 781}

\bibitem[\protect\citeauthoryear{{Comparat} et~al.,}{{Comparat}
  et~al.}{2016}]{Comparat:2016}
{Comparat} J.,  et~al., 2016, \mn@doi [\aap] {10.1051/0004-6361/201527377},
  \href {http://adsabs.harvard.edu/abs/2016A%26A...592A.121C} {592, A121}

\bibitem[\protect\citeauthoryear{{Contreras}, {Zehavi}, {Baugh}, {Padilla}  \&
  {Norberg}}{{Contreras} et~al.}{2017a}]{C17}
{Contreras} S.,  {Zehavi} I.,  {Baugh} C.~M.,  {Padilla} N.,   {Norberg} P.,
  2017a, \mn@doi [\mnras] {10.1093/mnras/stw2826}, \href
  {http://adsabs.harvard.edu/abs/2017MNRAS.465.2833C} {465, 2833}

\bibitem[\protect\citeauthoryear{{Contreras}, {Padilla}  \&
  {Lagos}}{{Contreras} et~al.}{2017b}]{C17b}
{Contreras} S.,  {Padilla} N.,   {Lagos} C.~D.~P.,  2017b, \mn@doi [\mnras]
  {10.1093/mnras/stx2410}, \href
  {http://adsabs.harvard.edu/abs/2017MNRAS.472.4992C} {472, 4992}

\bibitem[\protect\citeauthoryear{{Contreras}, {Zehavi}, {Padilla}, {Baugh},
  {Jim{\'e}nez}  \& {Lacerna}}{{Contreras} et~al.}{2018}]{C18}
{Contreras} S.,  {Zehavi} I.,  {Padilla} N.,  {Baugh} C.,  {Jim{\'e}nez} E.,
  {Lacerna} I.,  2018, preprint, \href
  {http://adsabs.harvard.edu/abs/2018arXiv180802896C} {} (\mn@eprint {arXiv}
  {1808.02896})

\bibitem[\protect\citeauthoryear{{Croton} et~al.,}{{Croton}
  et~al.}{2006}]{Croton:2006}
{Croton} D.~J.,  et~al., 2006, \mn@doi [\mnras]
  {10.1111/j.1365-2966.2005.09675.x}, \href
  {http://adsabs.harvard.edu/abs/2006MNRAS.365...11C} {365, 11}

\bibitem[\protect\citeauthoryear{{Croton}, {Gao}  \& {White}}{{Croton}
  et~al.}{2007}]{Croton:2007}
{Croton} D.~J.,  {Gao} L.,   {White} S.~D.~M.,  2007, \mn@doi [\mnras]
  {10.1111/j.1365-2966.2006.11230.x}, \href
  {http://adsabs.harvard.edu/abs/2007MNRAS.374.1303C} {374, 1303}

\bibitem[\protect\citeauthoryear{{Davis}, {Efstathiou}, {Frenk}  \&
  {White}}{{Davis} et~al.}{1985}]{Davis:1985}
{Davis} M.,  {Efstathiou} G.,  {Frenk} C.~S.,   {White} S. D.~M.,  1985, \apj,
  \href {http://adsabs.harvard.edu/abs/1985ApJ...292..371D} {292, 371}

\bibitem[\protect\citeauthoryear{{Dawson} et~al.,}{{Dawson}
  et~al.}{2016}]{Dawson:2016}
{Dawson} K.~S.,  et~al., 2016, \mn@doi [\aj] {10.3847/0004-6256/151/2/44},
  \href {http://adsabs.harvard.edu/abs/2016AJ....151...44D} {151, 44}

\bibitem[\protect\citeauthoryear{{De Lucia} \& {Blaizot}}{{De Lucia} \&
  {Blaizot}}{2007}]{DeLucia:2007}
{De Lucia} G.,  {Blaizot} J.,  2007, \mn@doi [\mnras]
  {10.1111/j.1365-2966.2006.11287.x}, \href
  {http://adsabs.harvard.edu/abs/2007MNRAS.375....2D} {375, 2}

\bibitem[\protect\citeauthoryear{{De Lucia}, {Kauffmann}  \& {White}}{{De
  Lucia} et~al.}{2004}]{DeLucia:2004}
{De Lucia} G.,  {Kauffmann} G.,   {White} S.~D.~M.,  2004, \mn@doi [\mnras]
  {10.1111/j.1365-2966.2004.07584.x}, \href
  {http://adsabs.harvard.edu/abs/2004MNRAS.349.1101D} {349, 1101}

\bibitem[\protect\citeauthoryear{{Eisenstein} et~al.,}{{Eisenstein}
  et~al.}{2011}]{Eisenstein:2011}
{Eisenstein} D.~J.,  et~al., 2011, \mn@doi [\aj] {10.1088/0004-6256/142/3/72},
  \href {http://adsabs.harvard.edu/abs/2011AJ....142...72E} {142, 72}

\bibitem[\protect\citeauthoryear{{Fisher}, {Davis}, {Strauss}, {Yahil}  \&
  {Huchra}}{{Fisher} et~al.}{1994}]{Fisher:1994}
{Fisher} K.~B.,  {Davis} M.,  {Strauss} M.~A.,  {Yahil} A.,   {Huchra} J.~P.,
  1994, \mn@doi [\mnras] {10.1093/mnras/267.4.927}, \href
  {http://adsabs.harvard.edu/abs/1994MNRAS.267..927F} {267, 927}

\bibitem[\protect\citeauthoryear{{Forero-Romero}, {Contreras}  \&
  {Padilla}}{{Forero-Romero} et~al.}{2014}]{FR:2014}
{Forero-Romero} J.~E.,  {Contreras} S.,   {Padilla} N.,  2014, \mn@doi [\mnras]
  {10.1093/mnras/stu1150}, \href
  {http://adsabs.harvard.edu/abs/2014MNRAS.443.1090F} {443, 1090}

\bibitem[\protect\citeauthoryear{{Gao} \& {White}}{{Gao} \&
  {White}}{2007}]{Gao:2007}
{Gao} L.,  {White} S.~D.~M.,  2007, \mn@doi [\mnras]
  {10.1111/j.1745-3933.2007.00292.x}, \href
  {http://adsabs.harvard.edu/abs/2007MNRAS.377L...5G} {377, L5}

\bibitem[\protect\citeauthoryear{{Gao}, {White}, {Jenkins}, {Stoehr}  \&
  {Springel}}{{Gao} et~al.}{2004}]{Gao:2004}
{Gao} L.,  {White} S.~D.~M.,  {Jenkins} A.,  {Stoehr} F.,   {Springel} V.,
  2004, \mn@doi [\mnras] {10.1111/j.1365-2966.2004.08360.x}, \href
  {http://adsabs.harvard.edu/abs/2004MNRAS.355..819G} {355, 819}

\bibitem[\protect\citeauthoryear{{Gao}, {Springel}  \& {White}}{{Gao}
  et~al.}{2005}]{Gao:2005}
{Gao} L.,  {Springel} V.,   {White} S.~D.~M.,  2005, \mn@doi [\mnras]
  {10.1111/j.1745-3933.2005.00084.x}, \href
  {http://adsabs.harvard.edu/abs/2005MNRAS.363L..66G} {363, L66}

\bibitem[\protect\citeauthoryear{{Guo} et~al.,}{{Guo} et~al.}{2011}]{Guo:2011}
{Guo} Q.,  et~al., 2011, \mn@doi [\mnras] {10.1111/j.1365-2966.2010.18114.x},
  \href {http://adsabs.harvard.edu/abs/2010arXiv1006.0106G} {413, 101}

\bibitem[\protect\citeauthoryear{{Guo}, {White}, {Angulo}, {Henriques},
  {Lemson}, {Boylan-Kolchin}, {Thomas}  \& {Short}}{{Guo}
  et~al.}{2013}]{Guo:2013a}
{Guo} Q.,  {White} S.,  {Angulo} R.~E.,  {Henriques} B.,  {Lemson} G.,
  {Boylan-Kolchin} M.,  {Thomas} P.,   {Short} C.,  2013, \mn@doi [\mnras]
  {10.1093/mnras/sts115}, \href
  {http://adsabs.harvard.edu/abs/2013MNRAS.428.1351G} {428, 1351}

\bibitem[\protect\citeauthoryear{{Guo} et~al.,}{{Guo} et~al.}{2015}]{Guo:2015a}
{Guo} H.,  et~al., 2015, \mn@doi [\mnras] {10.1093/mnras/stu2120}, \href
  {http://adsabs.harvard.edu/abs/2015MNRAS.446..578G} {446, 578}

\bibitem[\protect\citeauthoryear{{Guo} et~al.,}{{Guo} et~al.}{2016}]{Guo:2016b}
{Guo} Q.,  et~al., 2016, \mn@doi [\mnras] {10.1093/mnras/stw1525}, \href
  {http://adsabs.harvard.edu/abs/2016MNRAS.461.3457G} {461, 3457}

\bibitem[\protect\citeauthoryear{{Guzzo} et~al.,}{{Guzzo}
  et~al.}{2008}]{Guzzo:2008}
{Guzzo} L.,  et~al., 2008, \mn@doi [\nat] {10.1038/nature06555}, \href
  {http://adsabs.harvard.edu/abs/2008Natur.451..541G} {451, 541}

\bibitem[\protect\citeauthoryear{{Hamilton}}{{Hamilton}}{1992}]{Hamilton:1992}
{Hamilton} A.~J.~S.,  1992, \mn@doi [\apjl] {10.1086/186264}, \href
  {http://adsabs.harvard.edu/abs/1992ApJ...385L...5H} {385, L5}

\bibitem[\protect\citeauthoryear{{Hearin}}{{Hearin}}{2015}]{Hearin:2015}
{Hearin} A.~P.,  2015, \mn@doi [\mnras] {10.1093/mnrasl/slv064}, \href
  {http://adsabs.harvard.edu/abs/2015MNRAS.451L..45H} {451, L45}

\bibitem[\protect\citeauthoryear{{Henriques}, {White}, {Thomas}, {Angulo},
  {Guo}, {Lemson}  \& {Springel}}{{Henriques} et~al.}{2013}]{Henriques:2013}
{Henriques} B.~M.~B.,  {White} S.~D.~M.,  {Thomas} P.~A.,  {Angulo} R.~E.,
  {Guo} Q.,  {Lemson} G.,   {Springel} V.,  2013, \mn@doi [\mnras]
  {10.1093/mnras/stt415}, \href
  {http://adsabs.harvard.edu/abs/2013MNRAS.431.3373H} {431, 3373}

\bibitem[\protect\citeauthoryear{{Henriques}, {White}, {Thomas}, {Angulo},
  {Guo}, {Lemson}, {Springel}  \& {Overzier}}{{Henriques}
  et~al.}{2015}]{Henriques:2015}
{Henriques} B.~M.~B.,  {White} S.~D.~M.,  {Thomas} P.~A.,  {Angulo} R.,  {Guo}
  Q.,  {Lemson} G.,  {Springel} V.,   {Overzier} R.,  2015, \mn@doi [\mnras]
  {10.1093/mnras/stv705}, \href
  {http://adsabs.harvard.edu/abs/2015MNRAS.451.2663H} {451, 2663}

\bibitem[\protect\citeauthoryear{{Hu} \& {Haiman}}{{Hu} \&
  {Haiman}}{2003}]{Hu:2003}
{Hu} W.,  {Haiman} Z.,  2003, \mn@doi [\prd] {10.1103/PhysRevD.68.063004},
  \href {http://adsabs.harvard.edu/abs/2003PhRvD..68f3004H} {68, 063004}

\bibitem[\protect\citeauthoryear{{Kaiser}}{{Kaiser}}{1986}]{Kaiser:1986}
{Kaiser} N.,  1986, \mnras, \href
  {http://adsabs.harvard.edu/abs/1986MNRAS.222..323K} {222, 323}

\bibitem[\protect\citeauthoryear{{Kazin}, {S{\'a}nchez}  \& {Blanton}}{{Kazin}
  et~al.}{2012}]{Kazin:2012}
{Kazin} E.~A.,  {S{\'a}nchez} A.~G.,   {Blanton} M.~R.,  2012, \mn@doi [\mnras]
  {10.1111/j.1365-2966.2011.19962.x}, \href
  {http://adsabs.harvard.edu/abs/2012MNRAS.419.3223K} {419, 3223}

\bibitem[\protect\citeauthoryear{{Lacerna}, {Contreras}, {Gonz{\'a}lez},
  {Padilla}  \& {Gonzalez-Perez}}{{Lacerna} et~al.}{2018}]{Lacerna:2018}
{Lacerna} I.,  {Contreras} S.,  {Gonz{\'a}lez} R.~E.,  {Padilla} N.,
  {Gonzalez-Perez} V.,  2018, \mn@doi [\mnras] {10.1093/mnras/stx3253}, \href
  {http://adsabs.harvard.edu/abs/2018MNRAS.475.1177L} {475, 1177}

\bibitem[\protect\citeauthoryear{{Leja} et~al.,}{{Leja}
  et~al.}{2013}]{Leja:2013}
{Leja} J.,  et~al., 2013, \mn@doi [\apjl] {10.1088/2041-8205/778/2/L24}, \href
  {http://adsabs.harvard.edu/abs/2013ApJ...778L..24L} {778, L24}

\bibitem[\protect\citeauthoryear{{Mao}, {Zentner}  \& {Wechsler}}{{Mao}
  et~al.}{2017}]{Mao:2017}
{Mao} Y.-Y.,  {Zentner} A.~R.,   {Wechsler} R.~H.,  2017, preprint, \href
  {http://adsabs.harvard.edu/abs/2017arXiv170503888M} {} (\mn@eprint {arXiv}
  {1705.03888})

\bibitem[\protect\citeauthoryear{{McEwen} \& {Weinberg}}{{McEwen} \&
  {Weinberg}}{2018}]{Mcewen:2018}
{McEwen} J.~E.,  {Weinberg} D.~H.,  2018, \mn@doi [\mnras]
  {10.1093/mnras/sty882}, \href
  {http://adsabs.harvard.edu/abs/2018MNRAS.tmp..857M} {}

\bibitem[\protect\citeauthoryear{{Mundy}, {Conselice}  \& {Ownsworth}}{{Mundy}
  et~al.}{2015}]{Mundy:2015}
{Mundy} C.~J.,  {Conselice} C.~J.,   {Ownsworth} J.~R.,  2015, \mn@doi [\mnras]
  {10.1093/mnras/stv860}, \href
  {http://adsabs.harvard.edu/abs/2015MNRAS.450.3696M} {450, 3696}

\bibitem[\protect\citeauthoryear{{Okumura}, {Matsubara}, {Eisenstein}, {Kayo},
  {Hikage}, {Szalay}  \& {Schneider}}{{Okumura} et~al.}{2008}]{Okumura:2008}
{Okumura} T.,  {Matsubara} T.,  {Eisenstein} D.~J.,  {Kayo} I.,  {Hikage} C.,
  {Szalay} A.~S.,   {Schneider} D.~P.,  2008, \mn@doi [\apj] {10.1086/528951},
  \href {http://adsabs.harvard.edu/abs/2008ApJ...676..889O} {676, 889}

\bibitem[\protect\citeauthoryear{{Padilla}, {Christlein}, {Gawiser},
  {Gonz{\'a}lez}, {Guaita}  \& {Infante}}{{Padilla}
  et~al.}{2010}]{Padilla:2010}
{Padilla} N.~D.,  {Christlein} D.,  {Gawiser} E.,  {Gonz{\'a}lez} R.~E.,
  {Guaita} L.,   {Infante} L.,  2010, \mn@doi [\mnras]
  {10.1111/j.1365-2966.2010.17317.x}, \href
  {http://adsabs.harvard.edu/abs/2010MNRAS.409..184P} {409, 184}

\bibitem[\protect\citeauthoryear{{Padmanabhan} \& {White}}{{Padmanabhan} \&
  {White}}{2008}]{Padmanabhan:2008}
{Padmanabhan} N.,  {White} M.,  2008, \mn@doi [\prd]
  {10.1103/PhysRevD.77.123540}, \href
  {http://adsabs.harvard.edu/abs/2008PhRvD..77l3540P} {77, 123540}

\bibitem[\protect\citeauthoryear{{Peacock}}{{Peacock}}{1999}]{Peacock:1999}
{Peacock} J.~A.,  1999, {Cosmological Physics}

\bibitem[\protect\citeauthoryear{{Peebles}}{{Peebles}}{1980}]{Peebles:1980}
{Peebles} P.~J.~E.,  1980, {The large-scale structure of the universe}

\bibitem[\protect\citeauthoryear{{Seo}, {Eisenstein}  \& {Zehavi}}{{Seo}
  et~al.}{2008}]{Seo:2008}
{Seo} H.-J.,  {Eisenstein} D.~J.,   {Zehavi} I.,  2008, \mn@doi [\apj]
  {10.1086/527553}, \href {http://adsabs.harvard.edu/abs/2008ApJ...681..998S}
  {681, 998}

\bibitem[\protect\citeauthoryear{{Sheth} \& {Tormen}}{{Sheth} \&
  {Tormen}}{2004}]{Sheth:2004}
{Sheth} R.~K.,  {Tormen} G.,  2004, \mn@doi [\mnras]
  {10.1111/j.1365-2966.2004.07733.x}, \href
  {http://adsabs.harvard.edu/abs/2004MNRAS.350.1385S} {350, 1385}

\bibitem[\protect\citeauthoryear{{Shoji}, {Jeong}  \& {Komatsu}}{{Shoji}
  et~al.}{2009}]{Shoji:2009}
{Shoji} M.,  {Jeong} D.,   {Komatsu} E.,  2009, \mn@doi [\apj]
  {10.1088/0004-637X/693/2/1404}, \href
  {http://adsabs.harvard.edu/abs/2009ApJ...693.1404S} {693, 1404}

\bibitem[\protect\citeauthoryear{{Sin}, {Lilly}  \& {Henriques}}{{Sin}
  et~al.}{2017}]{Sin:2017}
{Sin} L.~P.~T.,  {Lilly} S.~J.,   {Henriques} B.~M.~B.,  2017, preprint, \href
  {http://adsabs.harvard.edu/abs/2017arXiv170208460S} {} (\mn@eprint {arXiv}
  {1702.08460})

\bibitem[\protect\citeauthoryear{{Springel}, {White}, {Tormen}  \&
  {Kauffmann}}{{Springel} et~al.}{2001}]{Springel:2001}
{Springel} V.,  {White} S. D.~M.,  {Tormen} G.,   {Kauffmann} G.,  2001,
  \mn@doi [\mnras] {10.1046/j.1365-8711.2001.04912.x}, \href
  {http://adsabs.harvard.edu/abs/2001MNRAS.328..726S} {328, 726}

\bibitem[\protect\citeauthoryear{{Springel} et~al.,}{{Springel}
  et~al.}{2005}]{Springel:2005}
{Springel} V.,  et~al., 2005, \mn@doi [\nat] {10.1038/nature03597}, \href
  {http://adsabs.harvard.edu/abs/2005Natur.435..629S} {435, 629}

\bibitem[\protect\citeauthoryear{{Strauss} et~al.,}{{Strauss}
  et~al.}{2002}]{Strauss:2002}
{Strauss} M.~A.,  et~al., 2002, \mn@doi [\aj] {10.1086/342343}, \href
  {http://adsabs.harvard.edu/abs/2002AJ....124.1810S} {124, 1810}

\bibitem[\protect\citeauthoryear{{Tinker}, {Hahn}, {Mao}, {Wetzel}  \&
  {Conroy}}{{Tinker} et~al.}{2018}]{Tinker:2018}
{Tinker} J.~L.,  {Hahn} C.,  {Mao} Y.-Y.,  {Wetzel} A.~R.,   {Conroy} C.,
  2018, \mn@doi [\mnras] {10.1093/mnras/sty666}, \href
  {http://adsabs.harvard.edu/abs/2018MNRAS.477..935T} {477, 935}

\bibitem[\protect\citeauthoryear{{Torrey} et~al.,}{{Torrey}
  et~al.}{2015}]{Torrey:2015}
{Torrey} P.,  et~al., 2015, \mn@doi [\mnras] {10.1093/mnras/stv1986}, \href
  {http://adsabs.harvard.edu/abs/2015MNRAS.454.2770T} {454, 2770}

\bibitem[\protect\citeauthoryear{{Wagner}, {M{\"u}ller}  \&
  {Steinmetz}}{{Wagner} et~al.}{2008}]{Wagner:2008}
{Wagner} C.,  {M{\"u}ller} V.,   {Steinmetz} M.,  2008, \mn@doi [\aap]
  {10.1051/0004-6361:20077688}, \href
  {http://adsabs.harvard.edu/abs/2008A%26A...487...63W} {487, 63}

\bibitem[\protect\citeauthoryear{{Wechsler}, {Zentner}, {Bullock}, {Kravtsov}
  \& {Allgood}}{{Wechsler} et~al.}{2006}]{Wechsler:2006}
{Wechsler} R.~H.,  {Zentner} A.~R.,  {Bullock} J.~S.,  {Kravtsov} A.~V.,
  {Allgood} B.,  2006, \mn@doi [\apj] {10.1086/507120}, \href
  {http://adsabs.harvard.edu/abs/2006ApJ...652...71W} {652, 71}

\bibitem[\protect\citeauthoryear{{Xu} \& {Zheng}}{{Xu} \&
  {Zheng}}{2018}]{Xu:2018}
{Xu} X.,  {Zheng} Z.,  2018, \mn@doi [\mnras] {10.1093/mnras/sty1547}, \href
  {http://adsabs.harvard.edu/abs/2018MNRAS.479.1579X} {479, 1579}

\bibitem[\protect\citeauthoryear{{Ye}, {Guo}, {Zheng}  \& {Zehavi}}{{Ye}
  et~al.}{2017}]{Ye:2017}
{Ye} J.-N.,  {Guo} H.,  {Zheng} Z.,   {Zehavi} I.,  2017, \mn@doi [\apj]
  {10.3847/1538-4357/aa70e7}, \href
  {http://adsabs.harvard.edu/abs/2017ApJ...841...45Y} {841, 45}

\bibitem[\protect\citeauthoryear{{Zehavi}, {Contreras}, {Padilla}, {Smith},
  {Baugh}  \& {Norberg}}{{Zehavi} et~al.}{2018}]{Zehavi:2018}
{Zehavi} I.,  {Contreras} S.,  {Padilla} N.,  {Smith} N.~J.,  {Baugh} C.~M.,
  {Norberg} P.,  2018, \mn@doi [\apj] {10.3847/1538-4357/aaa54a}, \href
  {http://adsabs.harvard.edu/abs/2018ApJ...853...84Z} {853, 84}

\bibitem[\protect\citeauthoryear{{Zentner}, {Hearin}  \& {van den
  Bosch}}{{Zentner} et~al.}{2014}]{Zentner:2014}
{Zentner} A.~R.,  {Hearin} A.~P.,   {van den Bosch} F.~C.,  2014, \mn@doi
  [\mnras] {10.1093/mnras/stu1383}, \href
  {http://adsabs.harvard.edu/abs/2014MNRAS.443.3044Z} {443, 3044}

\bibitem[\protect\citeauthoryear{{Zu} \& {Mandelbaum}}{{Zu} \&
  {Mandelbaum}}{2016}]{Zu:2016a}
{Zu} Y.,  {Mandelbaum} R.,  2016, \mn@doi [\mnras] {10.1093/mnras/stw221},
  \href {http://adsabs.harvard.edu/abs/2016MNRAS.457.4360Z} {457, 4360}

\bibitem[\protect\citeauthoryear{{Zu}, {Zheng}, {Zhu}  \& {Jing}}{{Zu}
  et~al.}{2008}]{Zu:2008}
{Zu} Y.,  {Zheng} Z.,  {Zhu} G.,   {Jing} Y.~P.,  2008, \mn@doi [\apj]
  {10.1086/591071}, \href {http://adsabs.harvard.edu/abs/2008ApJ...686...41Z}
  {686, 41}

\makeatother
\end{thebibliography}

\appendix
\section{Additional results}

\begin{figure*}
\includegraphics[width=0.47\textwidth]{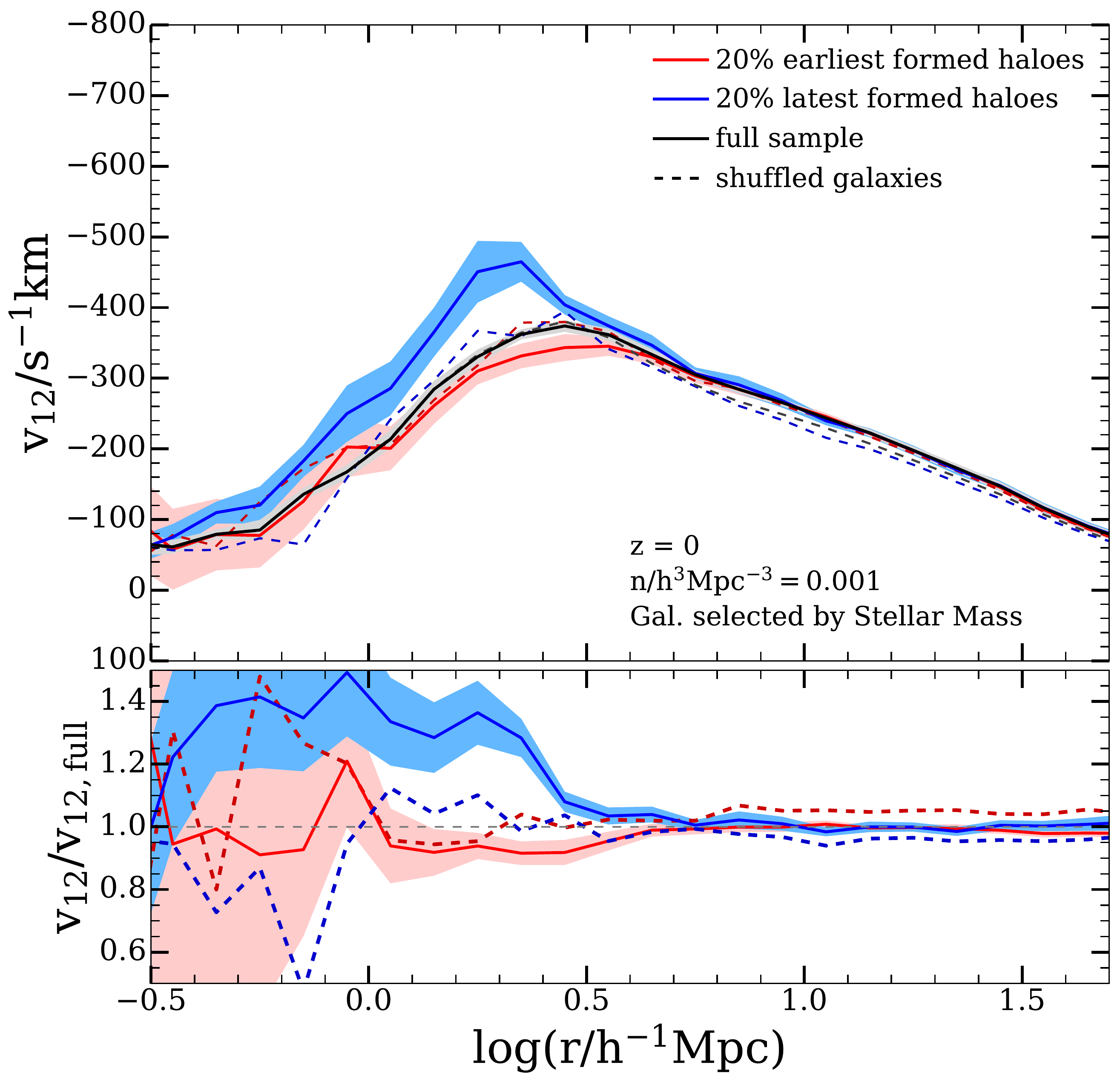}
\includegraphics[width=0.47\textwidth]{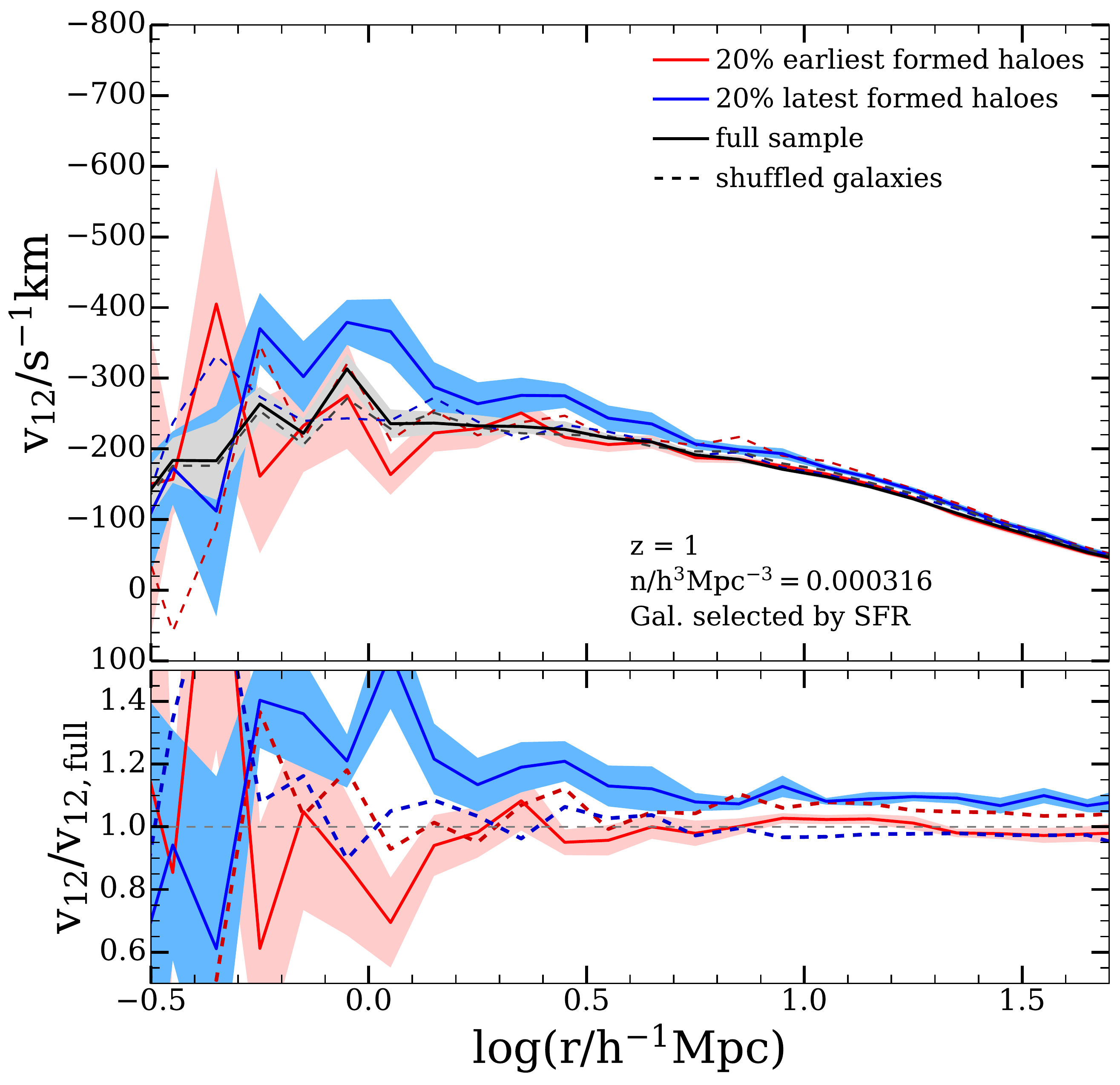}
\includegraphics[width=0.47\textwidth]{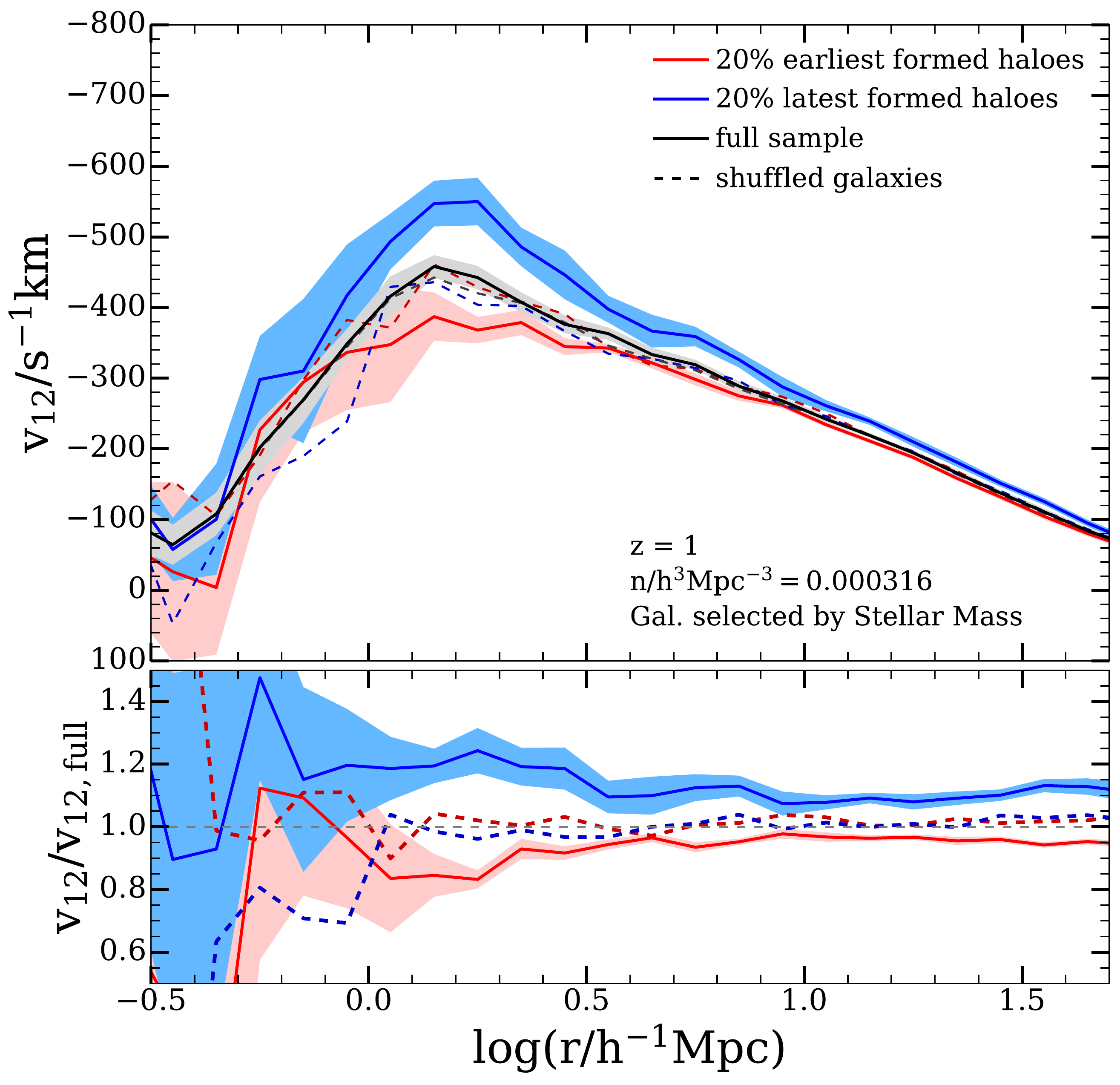}
\includegraphics[width=0.47\textwidth]{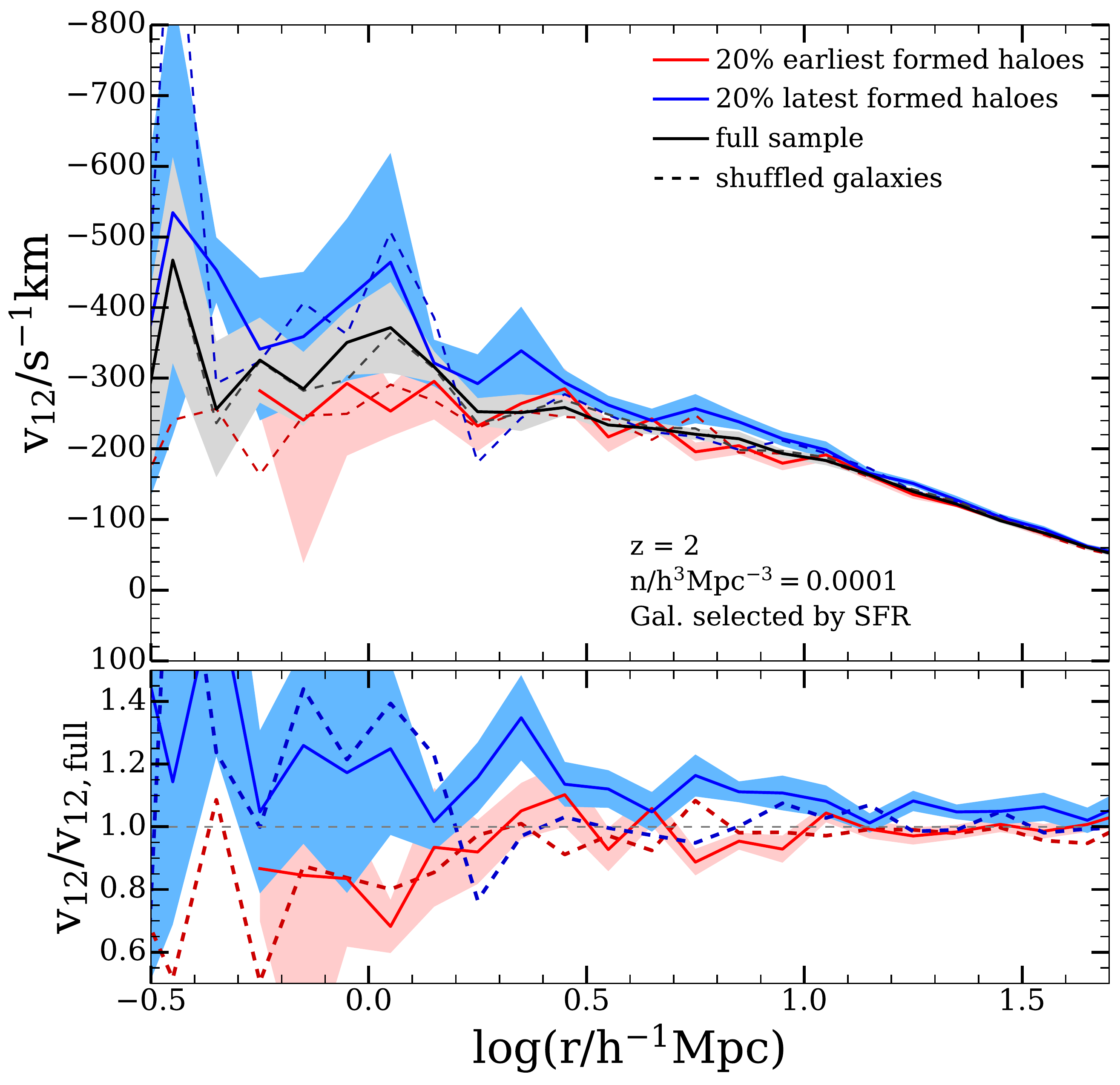}
\caption{Pairwise velocities for all galaxies in samples selected by stellar mass for $z=0$ and $z=1$ (top and bottom left panels, respectively), and galaxies selected by star formation rate for $z=1$ and $z=2$ (top and bottom right panels, respectively), with space densities indicated in the key.  The blue and
red lines correspond to the first and fifth quintiles in halo age.  The solid lines show the results for SAM galaxies in their original haloes, whereas the dashed lines are for the catalogues where galaxies where shuffled among haloes retaining the halo spatial and velocity structures}
\label{Fig:vcstm2}
\end{figure*}

\begin{figure}
\includegraphics[width=0.45\textwidth]{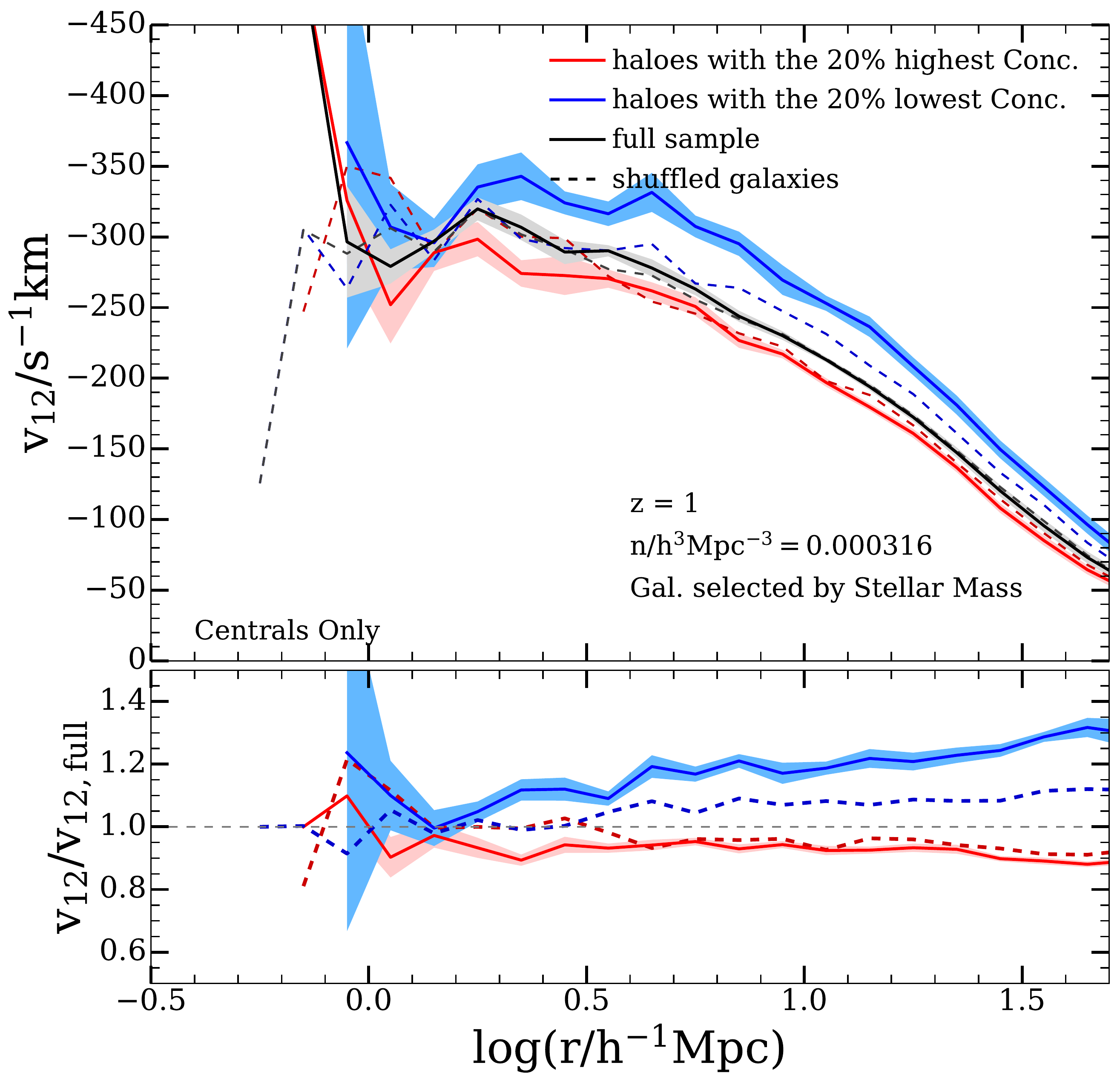}
\includegraphics[width=0.45\textwidth]{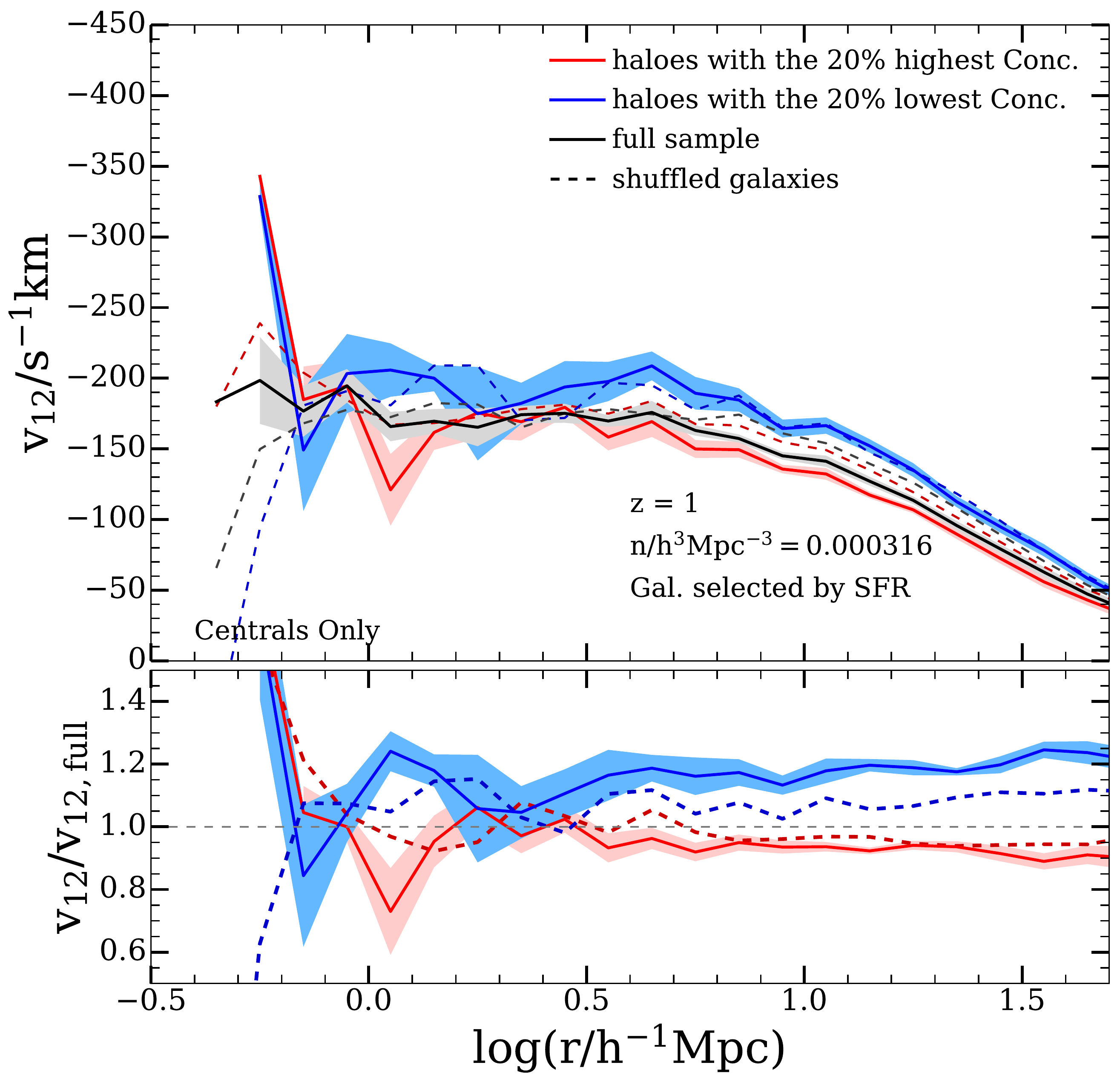}
\caption{Pairwise velocities for central galaxies in the samples for stellar mass (top) and star formation rate selection (bottom) for redshift $z=1$ and $2$ (top and bottom, respectively).  The case where only central galaxies are used shows a similar result for the 2-halo term, pointing to an assembly bias effect dominated by central pairs.}
\label{Fig:vccen}
\end{figure}

\begin{figure}
\includegraphics[width=0.44\textwidth]{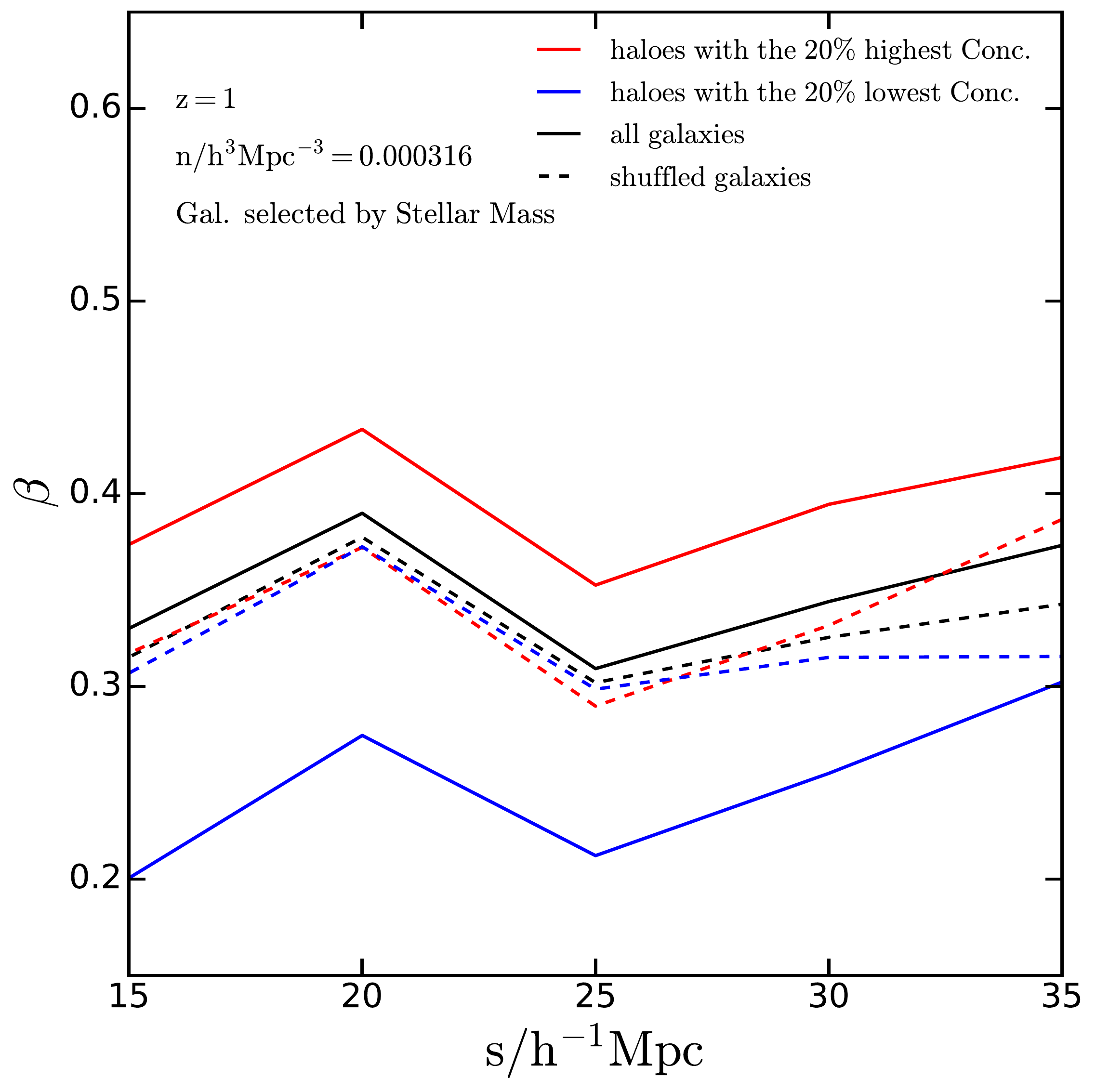}
\caption{Values of the $\beta$ parameter as a function of separation obtained from
the multipoles of the correlation function in redshift space. The results shown correspond a stellar mass selected sample with $n=0.000316\ {h}^3$Mpc$^{-3}$ (solid black line), and for the quintiles of high and
low halo concentration (solid red and blue lines, respectively), as a function of
separation.  The dashed lines are for the corresponding shuffled samples.}
\label{Fig:betamult}
\end{figure}

\begin{figure}
\includegraphics[width=0.45\textwidth]{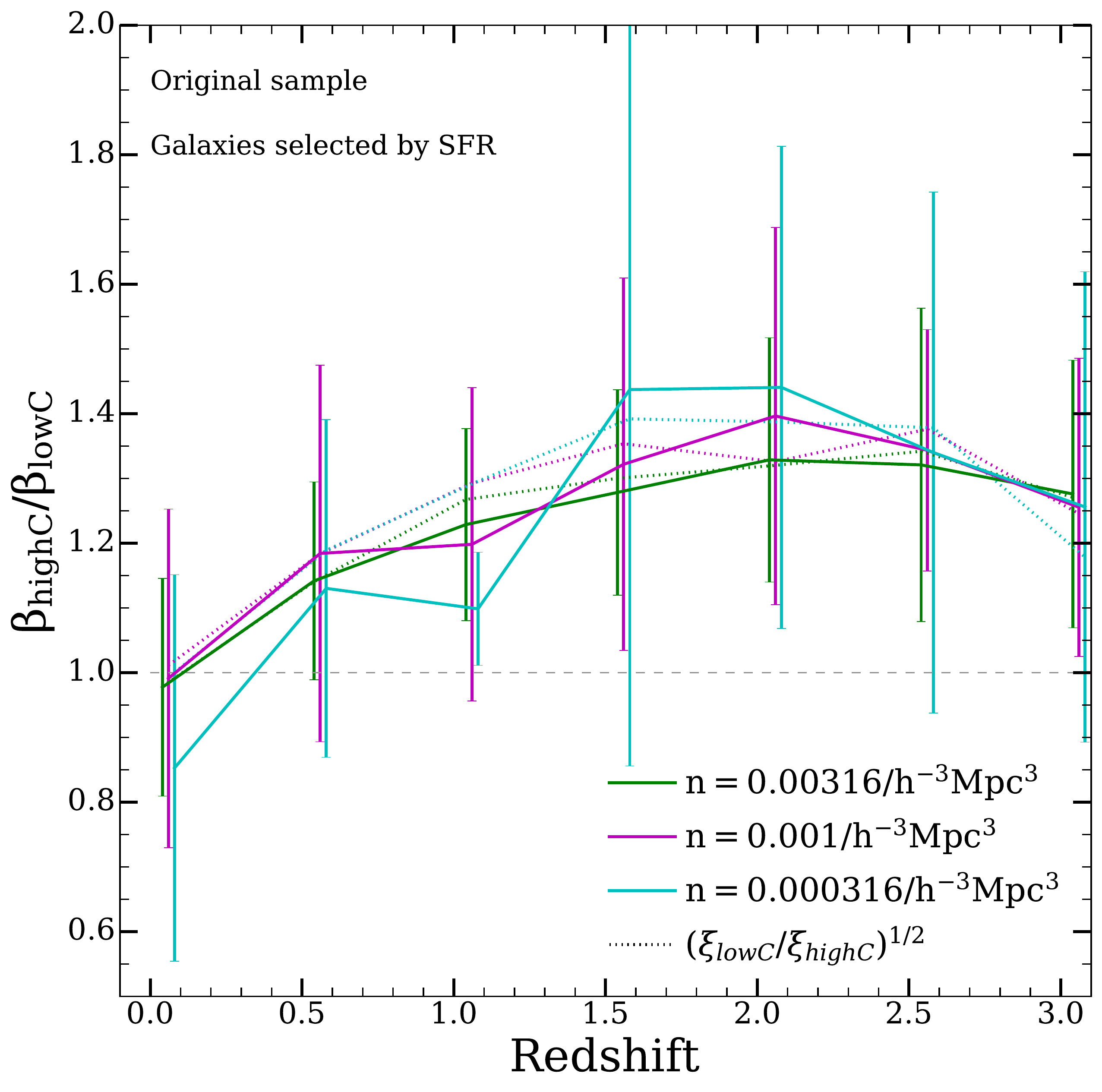}
\caption{Ratio between the averaged $\beta$ values for high and low concentration samples,
as a function of redshift, for different space density samples (colours, as indicated in the key)
selected by their star formation rate. Solid lines show results for the original samples; dashed lines
show the square root of the ratio between the real-space correlation functions for the same samples, averaged over the same range of separations; the agreement between solid and dashed lines shows that both velocities and clustering respond to assembly bias in similar ways.  The horizontal grey dashed line shows the unit ratio.  
Errors are obtained
from jackknife samples.}
\label{Fig:betaavsfr}
\end{figure}

In this Appendix we present several additional results that are mentioned in
the main text but which we felt would have distracted the reader from the flow
of the paper.

In the main text we concentrated our analysis on halo concentration
as the secondary halo parameter. Figure \ref{Fig:vcstm2} shows the results for the
pairwise velocities when using halo age as the secondary parameter.  In this
case the results indicate that halo assembly bias and occupancy variation 
roughly cancel one another in the $z=0$ sample, 
as evidenced by
a noticeable difference in the pairwise velocities of early and late formed
haloes for the shuffled samples, with younger haloes showing a lower amplitude of infall velocity.  In this case halo assembly bias increases the infall for early formed haloes, and occupancy variation lowers it for the same population.  

Differences in pairwise velocities are also present in the $z=1$ SFR selected samples of the top-right panel. At this redshift there is a net galaxy assembly bias effect on velocities, and again halo assembly bias imprints the opposite trend than occupancy variation.   At the same redshift, samples selected by stellar mass show only an occupancy variation effect (lower left).  The $z=2$ example on the lower right shows effects
from occupancy variation alone, although small, and no halo assembly bias.  These results are consistent with those found by \cite{C18} for the clustering
amplitudes of samples separated by halo age as secondary
property at different redshifts.

We show the pairwise velocities for different quintiles
of concentration for central galaxies only in Figure \ref{Fig:vccen}.
By comparing to Fig. \ref{Fig:vcstm} it is clear that the results remain unchanged, with similar amplitude differences for high and low concentration
samples selected by either stellar mass or SFR at $z=1$.  
The differences between
shuffled samples are also similar to when using the full population of galaxies.
This indicates that centrals dominate the amplitude of the pairwise 
velocities in our chosen samples, and that the shuffled catalogues simply
remove the dependence of 
the minimum halo mass to host a central galaxy on secondary halo parameters.  These results are consistent with \cite{Zehavi:2018}.

In the main text, we calculated the $\beta$ parameter using the ratio of the monopoles of the correlation functions in redshift and real space (see Fig. \ref{Fig:beta}).  Figure \ref{Fig:betamult} shows the results from combining Eqs. 26 and 27 of \cite{Hamilton:1992}, which is the method that can be adopted when analysing actual observational data that consist of galaxies with distances measured via their redshifts.  As can be seen the results are roughly consistent with those from the monopoles, with galaxies in the original sample with higher concentration showing a larger value of $\beta$ at all scales.  Qualitatively
this trend is also present in the shuffled samples, but the amplitude difference
is smaller; these differences should be revisited with a larger simulation in future work.

Figure \ref{Fig:betaavsfr} shows the ratio between $\beta$ values for samples in the upper and lower
concentration quintiles in SFR selected samples.  This confirms the
results already obtained for stellar mass selected samples (Fig. \ref{Fig:betaav}), that the change
in the $\beta$ parameter obtained from redshift-space distortions for
samples in the extreme quintiles of halo concentration simply respond
to their change in bias factor introduced by the split in concentration.  The figure is for the original
samples; the shuffled samples show equivalent results.

\label{lastpage}
\end{document}